\documentclass[reprint]{revtex4-2}
\usepackage{amsmath,amssymb,amsfonts,amsthm,enumitem,braket,bbm,mathtools,tikz}
\usepackage[hidelinks]{hyperref}

\newcommand{\R}{\mathbb{R}}
\newcommand{\Z}{\mathbb{Z}}
\newcommand{\bs}{\boldsymbol}
\theoremstyle{definition}

\DeclareMathOperator{\im}{im}

\begin{document}

\title{Stabilizer models of achiral abelian topological orders}
\author{Marvin Qi and Wilbur Shirley}
\affiliation{Leinweber Institute for Theoretical Physics, University of Chicago, Chicago, IL 60637, USA}
\date{\today}

\begin{abstract}
    We construct continuous-variable stabilizer codes realizing every achiral 2D abelian topological order, both bosonic and fermionic. Starting from an arbitrary signature-zero integer $K$-matrix, we show that the corresponding topological order can be obtained by condensing a lattice of dyons with trivial pairwise mutual braiding in an $\R^N$ gauge theory. We implement this condensation microscopically by gauging a non-on-site 1-form symmetry of the $\R^N$ toric code.
\end{abstract}

\maketitle

\section{Introduction}

Topological stabilizer codes provide exactly solvable models of abelian topological phases, allowing their universal properties to be understood directly at the microscopic level. The 2D toric code is the paradigmatic example: the fusion rules and braiding statistics of its anyonic excitations can be derived explicitly from the underlying lattice model~\cite{Kitaev2003Anyons}. At the same time, topological stabilizer codes are a promising avenue toward quantum error correction, since they encode quantum information nonlocally and are therefore intrinsically robust against local errors~\cite{Dennis2002Memory}.

In Ref.~\cite{Ellison2022TwistedQuantumDoubles}, it was shown that every \textit{gappable} 2D bosonic abelian topological order---\textit{i.e.}, one that admits a gapped boundary~\cite{Levin2013ProtectedEdge}---can be realized by a topological stabilizer code on a system of qudits. For example, the double semion theory~\cite{LevinWen2005StringNet} admits a stabilizer code realization using $\Z_4$ qudits~\cite{Ellison2022TwistedQuantumDoubles}. This result has been extended to the fermionic setting in Ref.~\cite{Sun2026MajoranaPauli}, which constructed Majorana-qudit stabilizer code models of all gappable fermionic abelian topological orders. Conversely, it is believed that 2D qudit or Majorana-qudit stabilizer codes can only realize gappable abelian topological orders; this has been proven in the translation-invariant bosonic setting~\cite{RubaYang2025Witt}. Thus, gappability is conjectured to precisely characterize the class of 2D abelian topological orders realizable by stabilizer codes on finite-dimensional local Hilbert spaces.

The authors of Ref.~\cite{Magdalena2026} constructed 2D continuous-variable (CV) stabilizer codes~\cite{Gottesman2001Oscillator} realizing certain \textit{achiral} topological orders that are nevertheless ungappable, such as that of $U(1)_2\times U(1)_{-4}$ Chern-Simons theory \cite{Levin2013ProtectedEdge}. These examples demonstrate that CV stabilizer codes can realize topological orders beyond those accessible to qudit stabilizer codes, raising a natural question: which 2D topological orders can be realized by CV stabilizer codes? In this paper, we construct stabilizer code realizations of all achiral abelian topological orders, both bosonic and fermionic. We further conjecture that achirality precisely characterizes the class of 2D abelian topological orders realizable by CV stabilizer codes.

Building on the insights of Ref.~\cite{Magdalena2026}, our construction begins from the observation that every achiral bosonic abelian topological order can be obtained by condensing a $2N$-dimensional lattice of mutually bosonic dyons in an $\R^N$ gauge theory. To implement this condensation in a quantum lattice model, we start from the $\R^N$ toric code and gauge the anomaly-free $\Z^{2N}$ 1-form symmetry generated by closed string operators of the mutually bosonic dyons. This yields a CV stabilizer code realizing the desired topological order. We further generalize this approach to achiral fermionic topological orders, which can be obtained by condensing fermionic lattices in $\R^N$ gauge theory. This requires the introduction of fermionic modes that trivialize the anomaly of the corresponding 1-form symmetry~\cite{Bhardwaj2017SpinTFT}. Gauging the $\Z^{2N}$ 1-form symmetry then yields a Majorana-CV stabilizer code realizing the achiral fermionic topological order.\\

\begin{figure*}[t]
    \begin{tikzpicture}[scale=.6]
        \pgfmathsetmacro{\z}{3.1}
        \pgfmathsetmacro{\sq}{sqrt(2)}
        \clip (-\z,-\z) rectangle (\z,\z);
        \draw (0,-\z)--(0,\z);
        \draw (-\z,0)--(\z,0);
        \foreach \x in {-3,...,3} {
            \foreach \y in {-3,...,3} {
            \fill (\sq*\x,\sq*\y) circle (2.8pt);
            }
        }
        \foreach \x in {-7,...,7} {
            \foreach \y in {-7,...,7} {
            \fill (.5*\sq*\x,.5*\sq*\y) circle (1.2pt);
            }
        }
    \end{tikzpicture}\qquad\quad
    \begin{tikzpicture}[scale=.6]
        \pgfmathsetmacro{\z}{3.1}
        \clip (-\z,-\z) rectangle (\z,\z);
        \draw (0,-\z)--(0,\z);
        \draw (-\z,0)--(\z,0);
        \foreach \x in {-5,...,5} {
            \foreach \y in {-5,...,5} {
            \fill (\x-\y,\x+\y) circle (2.8pt);
            }
        }
        \foreach \x in {-5,...,5} {
            \foreach \y in {-8,...,8} {
            \fill (.5*\x-.5*\y,.5*\x+.5*\y) circle (1.2pt);
            }
        }
    \end{tikzpicture}\qquad\quad
    \begin{tikzpicture}[scale=.6]
        \pgfmathsetmacro{\z}{3.1}
        \pgfmathsetmacro{\sq}{sqrt(2)}
        \pgfmathsetmacro{\sqf}{sqrt(2)/4}
        \clip (-\z,-\z) rectangle (\z,\z);
        \draw (0,-\z)--(0,\z);
        \draw (-\z,0)--(\z,0);
        \foreach \x in {-5,...,5} {
            \foreach \y in {-5,...,5} {
            \fill (\x-\sq*\y,\x+\sq*\y) circle (2.8pt);
            }
        }
        \foreach \x in {-5,...,5} {
            \foreach \y in {-8,...,8} {
            \fill (.5*\x-\sqf*\y,.5*\x+\sqf*\y) circle (1.2pt);
            }
        }
    \end{tikzpicture}\qquad\quad
    \begin{tikzpicture}[scale=.6]
        \pgfmathsetmacro{\z}{3.1}
        \pgfmathsetmacro{\ax}{(5+sqrt(5))/10}
        \pgfmathsetmacro{\ay}{(5-sqrt(5))/10}
        \pgfmathsetmacro{\bx}{1/sqrt(5)}
        \pgfmathsetmacro{\by}{-1/sqrt(5)}
        \pgfmathsetmacro{\cx}{1}
        \pgfmathsetmacro{\cy}{1}
        \pgfmathsetmacro{\dx}{-(1+sqrt(5))/2}
        \pgfmathsetmacro{\dy}{(sqrt(5)-1)/2}
        \clip (-\z,-\z) rectangle (\z,\z);
        \draw (0,-\z)--(0,\z);
        \draw (-\z,0)--(\z,0);
        \foreach \c in {-5,...,5} {
            \foreach \d in {-5,...,5} {
            \fill (\cx*\c+\dx*\d,\cy*\c+\dy*\d) circle (2.8pt);
            }
        }
        \foreach \a in {-5,...,5} {
            \foreach \b in {-8,...,8} {
            \fill (\ax*\a+\bx*\b,\ay*\a+\by*\b) circle (1.2pt);
            }
        }
    \end{tikzpicture}
    (a)\hspace{121.5pt}(b)\hspace{121.5pt}(c)\hspace{121.5pt}(d)
    \caption{Condensate lattice $L$ (large dots) and dual lattice $L^\vee$ (small dots) for (a) $\Z_2$ toric code, (b) double semion, (c) $U(1)_2\times U(1)_{-4}$, and (d) the achiral $\Z_5$ theory.}
    \label{fig:lattices}
\end{figure*}
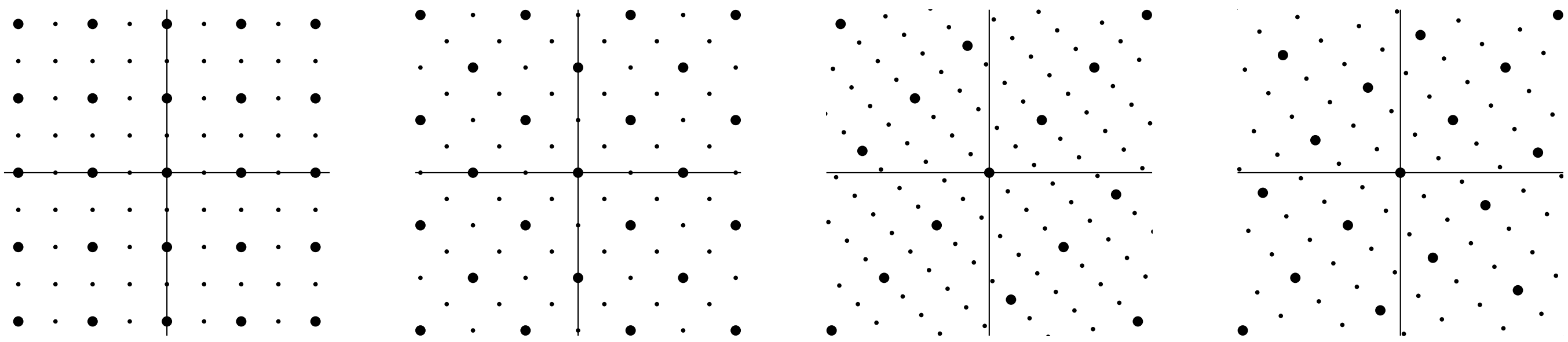

\section{Dyon condensation in $\R^N$ gauge theory}

We begin by reviewing the formalism of dyon condensation in $\R^N$ gauge theory~\cite{Magdalena2026}, allowing both electric charges and magnetic fluxes as dynamical excitations. The dyons of the theory are labeled by their flux and charge,
\begin{equation}
    (\bs\phi,\bs q)=(\phi^1,\ldots,\phi^N,q^1,\ldots,q^N)\in\R^{2N},
\end{equation}
with fusion given by addition in $\R^{2N}$. We use a normalization in which $\phi^n=1$ represents $2\pi$ flux of the $n$th gauge field. The topological spins and mutual braiding statistics are encoded by the symmetric bilinear form
\begin{equation}
    \label{eq:J}
    \langle x,y\rangle=x^TJy,\qquad J=\begin{pmatrix}0&I_N\\I_N&0\end{pmatrix}.
\end{equation}
In particular, the topological spin and braiding are
\begin{equation}
    \theta_x=e^{\pi i\langle x,x\rangle},\qquad B_{x,y}=e^{2\pi i\langle x,y\rangle},
    \label{eq:spinbraiding}
\end{equation}
for any pair of dyons $x,y\in\R^{2N}$.

Consider a $2N$-dimensional lattice of dyons $L\subset\R^{2N}$. 
Assume that the dyons in $L$ have trivial pairwise braiding, that is,
\begin{equation}
    \langle c,c'\rangle\in\Z,\qquad c,c'\in L.
    \label{eq:trivialmutual}
\end{equation}
In particular, $\theta_c=\pm1$, so every dyon is either a boson or a fermion. We call $L$ \textit{bosonic} if every dyon in $L$ is a boson, and \textit{fermionic} otherwise.

A bosonic (resp. fermionic) lattice can be condensed in a bosonic (resp. fermionic) system~\cite{Bais2009Condensation,Bhardwaj2017SpinTFT}. Condensing $L$ has two effects. First, dyons that braid nontrivially with any $c\in L$ become confined; the dyons that survive as deconfined excitations are those which braid trivially with $L$, and therefore form the dual lattice
\begin{equation}
    L^\vee=\{a\in\R^{2N}:\langle a,c\rangle\in\Z\,\,\,\forall c\in L\}.
\end{equation}
Note that $L\subseteq L^\vee$. Second, deconfined dyons that differ by fusion with an element of $L$ become identified. Thus, the fusion group of the resulting topological order, modulo local excitations, is the discriminant group~\cite{Nikulin1980Lattices}
\begin{equation}
    \mathcal{A}=L^\vee/L.
\end{equation}

We illustrate the formalism with four examples obtained by condensing two-dimensional dyon lattices in $\R$ gauge theory. The condensate lattice $L$ is generated by $c_1$ and $c_2$, and the dual lattice $L^\vee$ is generated by $a_1$ and $a_2$. These lattices are depicted in Fig.~\ref{fig:lattices}.\\

\noindent 1) $\Z_2$ toric code:
\begin{align*}
    c_1&=(0,2),&
    a_1&=\left(\tfrac12,0\right),\\ c_2&=(1,0),&
    a_2&=(0,1).
\end{align*}
$L$ is generated by charge $2$ and $2\pi$ flux. The unit charge $a_2$ and $\pi$ flux $a_1$ descend, respectively, to the $e$ and $m$ anyons of the toric code. Equivalently, we may take $c_1=(0,\sqrt{2})$, $c_2=(\sqrt{2},0)$, as shown in Fig.~\ref{fig:lattices}(a).\\

\noindent 2) Double semion:
\begin{align*}
    c_1&=(1,1),&
    a_1&=\left(\tfrac12,\tfrac12\right),\\ c_2&=(-1,1),&
    a_2&=\left(\tfrac12,-\tfrac12\right).
\end{align*}
Here, $[a_1]$ and $[a_2]$ are a semion and antisemion, respectively. This condensation can be viewed as a two-step process: first condense $(2,0)$ and $(0,2)$ to obtain the $\Z_4$ toric code, then condense its $e^2m^2$ boson, corresponding to $[c_1]=[c_2]$. This example admits the following generalization:

\noindent 3) $U(1)_m\times U(1)_{-n}$:
\begin{align*}
    c_1&=\left(\sqrt{\tfrac{m}{2}},\sqrt{\tfrac{m}{2}}\right),&
    a_1&=\left(\sqrt{\tfrac{1}{2m}},\sqrt{\tfrac{1}{2m}}\right),\\
    c_2&=\left(-\sqrt{\tfrac{n}{2}},\sqrt{\tfrac{n}{2}}\right),&
    a_2&=\left(\sqrt{\tfrac{1}{2n}},-\sqrt{\tfrac{1}{2n}}\right).
\end{align*}
The classes $[a_1]$ and $[a_2]$ generate the $U(1)_m$ and $U(1)_{-n}$ factors, respectively. The theory is bosonic when both $m$ and $n$ are even, and fermionic otherwise.\\

\noindent 4) Achiral $\Z_p$ theory, $p\equiv1\pmod 4$:
\begin{align*}
    c_1&=(1,1),&
    a_1&=\left(\tfrac{p-\sqrt p}{2p},\tfrac{p+\sqrt p}{2p}\right),\\
    c_2&=\left(\tfrac{1-\sqrt p}{2},\tfrac{1+\sqrt p}{2}\right),&
    a_2&=\left(\tfrac{\sqrt p}{p},-\tfrac{\sqrt p}{p}\right).
\end{align*}
The condensate lattice is bosonic, with discriminant group $\mathcal{A}\cong\Z_p$. It is generated by $[a_1]$, with $[a_2]=-2[a_1]$, and the generator has topological spin
\begin{equation}
    \theta_{[a_1]}=e^{2\pi i k/p},\qquad k=\frac{p-1}{4}.
    \label{eq:Zptopologicalspin}
\end{equation}

We note that these examples include minimal representatives of all nontrivial classes in the achiral subgroups of the prime components of the pointed Witt group~\cite{Davydov2013Witt}. Let $\mathcal{W}^\mathrm{ach}_\mathrm{pt}\subset\mathcal{W}_\mathrm{pt}$ denote the subgroup represented by achiral abelian bosonic topological orders, and define
\begin{equation}
    \mathcal{W}^\mathrm{ach}_\mathrm{pt}(p)
    :=\mathcal{W}_\mathrm{pt}(p)\cap\mathcal{W}^\mathrm{ach}_\mathrm{pt}
    \cong
    \begin{cases}
        \Z_2,&p=2,\\
        \Z_2,&p\equiv1\pmod4,\\
        0,&p\equiv3\pmod4.
    \end{cases}
\end{equation}
Minimal representatives of the nontrivial classes are given by $U(1)_2\times U(1)_{-4}$ for $p=2$ and the achiral $\Z_p$ theory for each prime $p\equiv1\pmod4$.\\

\section{Chern-Simons theory description}

Condensing the dyon lattice $L$ amounts to gauging the $1$-form symmetry generated by the corresponding Wilson lines, since gauging has the effect of proliferating symmetry defects~\cite{Gaiotto2015Generalized}---which are precisely the dyons in $L$~\cite{Gaiotto2015Generalized,Bhardwaj2017SpinTFT}. Here we describe this gauging in the continuum theory. As we will see, this gives rise to a $U(1)^{2N}$ Chern-Simons theory with $K$-matrix equal to the Gram matrix of $L$:
\begin{equation}
    K_{IJ}=\langle c_I,c_J\rangle,
    \label{eq:Gram}
\end{equation}
where
\begin{equation}
    c_I=(\bs\phi_I,\bs q_I),\qquad I=1,\ldots,2N,
\end{equation}
is a basis of $L$.

First, recall the BF theory~\cite{Horowitz1989BF} description of the topological sector of $\R^N$ gauge theory. The theory has $N$ noncompact gauge fields $\alpha_I$ and $N$ noncompact Lagrange multiplier fields $\beta_I$, with action
\begin{equation}
    S_\mathrm{BF}=-\frac{1}{2\pi}\int \beta_I\wedge d\alpha_I = -\frac{1}{4\pi} \int J_{IJ} B_I \wedge d B_J.
\end{equation}
Here we have defined the noncompact $\R^{2N}$ gauge field 
\begin{equation}
    B=(\bs\beta,\bs\alpha)=(\beta_1,\ldots,\beta_N,\alpha_1,\ldots,\alpha_N).
\end{equation}
The equations of motion impose the flatness condition $d B_I = 0$.
The action is invariant under $\R^{2N}$ gauge transformations 
\begin{equation}
    B_I \to B_I + d f_I.
\end{equation}
Since the gauge fields are noncompact, $f$ is a genuine $\R^{2N}$-valued function.
Wilson lines are labeled by arbitrary dyons $a = (\bs \phi, \bs q) \in\R^{2N}$ as
\begin{equation}
    W_a(\gamma) = \exp \left[ i \int_\gamma a_I B_I \right].
\end{equation}
The $\R^{2N}$ 1-form symmetry acts as $B_I \to B_I + J_{IJ} \Lambda_J$ for closed $\Lambda_I$; its action on Wilson lines is 
\begin{equation}
    W_a(\gamma) \to e^{2\pi i \langle a, \Gamma \rangle  }W_a(\gamma),
\end{equation}
where $\Gamma_I = \frac{1}{2\pi}\int_\gamma \Lambda_I$. Note that $\Gamma_I$ is unquantized, making this a $\R^{2N}$ 1-form symmetry transformation. We are interested in gauging $\Z^{2N}$ subgroups of this 1-form symmetry. 

Let $M = \begin{pmatrix}c_1&\cdots&c_{2N}\end{pmatrix}$ be the matrix of condensed dyons, such that the condensate lattice $L = M \Z^{2N}$. Consider the $\Z^{2N}$ 1-form subgroup consisting of transformations 
\begin{equation} \label{eq:Z1form}
    B_I \to B_I +  J_{IJ} M_{JK} \lambda_K,
\end{equation}
where $\lambda_K$ are closed ($d \lambda_K = 0$) and obey the quantization $\int_\gamma \lambda_K \in 2\pi\Z$. Gauging this 1-form symmetry identifies the gauge field configurations which differ by \eqref{eq:Z1form}. 
In other words, these transformations become gauge transformations. Requiring invariance of the Wilson lines $W_a(\gamma)$ imposes the quantization condition
\begin{equation}
M^T J a \in \Z^{2N}.
\end{equation}
We may therefore label these deconfined Wilson lines by an integer vector $\ell = M^T J a$. Defining $A = (JM)^{-1} B$, the deconfined Wilson lines are 
\begin{equation}
    W_a(\gamma) = \exp \left[ i \int_\gamma \ell_I A_I \right].
\end{equation}
Hence $A_I$ are properly quantized $U(1)$ gauge fields. Substituting $B=(JM)A$ into $S_\mathrm{BF}$ gives
\begin{equation} \label{eq:KmatrixCS}
    S=-\frac{1}{4\pi}\int K_{IJ}A_I\wedge dA_J.
\end{equation}
We thus obtain a $U(1)^{2N}$ Chern-Simons theory with $K$-matrix 
\begin{equation}
    K=M^TJM.
\end{equation}
The condition \eqref{eq:trivialmutual} follows from the requirement that the theory is properly quantized, $K_{IJ} \in \Z$, and therefore gauge invariant. The theory thus describes a 2D abelian topological order~\cite{Wen1992Classification}. 
Moreover, $K$ is even, meaning that $K_{II}\in2\Z$ for all $I$, precisely when $L$ is bosonic. (When $K$ is odd, the theory is a spin Chern–Simons theory, whose definition requires a choice of spin structure on spacetime~\cite{Belov2005SpinCS}.)

The anyons of the theory are labeled by integer vectors $\ell\in\Z^{2N}$. Vectors in $K\Z^{2N}$ represent local excitations, which are all bosonic when $K$ is even, but otherwise include fermions. The fusion group is thus
\begin{equation}
    \mathcal{A}=\Z^{2N}/K\Z^{2N}\cong L^\vee/L,
    \label{eq:iso}
\end{equation}
in agreement with the dyon condensation picture. To see the second isomorphism, observe that a dyon $a\in\R^{2N}$ belongs to $L^\vee$ precisely when $M^TJa\in\Z^{2N}$, and hence $L^\vee=(M^TJ)^{-1}\Z^{2N}$. Under the isomorphism $L^\vee\to\Z^{2N}$ given by $a\mapsto M^TJa$, the sublattice $L=M\Z^{2N}$ maps to $K\Z^{2N}$. This gives the isomorphism $L^\vee/L\cong\Z^{2N}/K\Z^{2N}$.

The topological spins and mutual braiding statistics are~\cite{Wen1992Classification,Belov2005SpinCS}
\begin{equation}
    \theta_\ell=e^{\pi i\ell^TK^{-1}\ell},\qquad
    B_{\ell,\ell'}=e^{2\pi i\ell^TK^{-1} \ell'}.
\end{equation}
For odd $K$, the topological spin depends on the representative of the equivalence class: attaching a local fermion reverses its sign. These expressions also agree with the dyon condensation picture, since the Gram matrix of $L^\vee$ is $K^{-1}$. Indeed, defining $a_I=(M^TJ)^{-1}\ell_I$, where $\ell_I$ is the $I$th basis vector of $\Z^{2N}$, gives
\begin{equation}
    (K^{-1})_{IJ}=\langle a_I,a_J\rangle.
\end{equation}
For instance, the lattice $L$ of the above example 4 gives
\begin{equation}
    K=\begin{pmatrix}2&1\\1&\frac{1-p}{2}\end{pmatrix},\qquad
    K^{-1}=\frac{1}{p}\begin{pmatrix}\frac{p-1}{2}&1\\1&-2\end{pmatrix}.
\end{equation}
Under the isomorphism in (\ref{eq:iso}), the anyon $[a_1]\in L^\vee/L$ corresponds to $[\ell_1]\in\Z^{2N}/K\Z^{2N}$.

Finally, the chiral central charge is given by the signature of $K$~\cite{Wen1995Edge},
\begin{equation}
    c_-=\operatorname{sig}(K)=n_+-n_-,
\end{equation}
where $n_\pm$ denote the numbers of positive and negative eigenvalues of $K$. Since $K$ is congruent to $J$, it has the same signature by Sylvester's law of inertia. As $\operatorname{sig}(J)=0$, the resulting topological order is achiral, with $c_-=0$.

Conversely, every achiral $U(1)^{2N}$ Chern-Simons theory arises from dyon condensation in $\R^N$ gauge theory. To see this, diagonalize $K$ as
\begin{equation}
    K=O^T\begin{pmatrix}D_+&0\\0&-D_-\end{pmatrix}O,
\end{equation}
where $O$ is orthogonal and $D_\pm$ are positive diagonal $N\times N$ matrices. Define
\begin{equation}
    M=H\begin{pmatrix}D_+^{1/2}&0\\0&D_-^{1/2}\end{pmatrix}O,
    \qquad
    H=\frac{1}{\sqrt2}\begin{pmatrix}I_N&I_N\\I_N&-I_N\end{pmatrix}.
\end{equation}
Since
\begin{equation}
    H^TJH=\begin{pmatrix}I_N&0\\0&-I_N\end{pmatrix},
\end{equation}
the lattice $L=M\Z^{2N}$ satisfies $K=M^TJM$. As every achiral 2D abelian topological order can be described by a zero-signature $K$-matrix~\cite{Wen1992Classification,Belov2005SpinCS}, we conclude that dyon condensation in $\R^N$ gauge theory realizes precisely the class of achiral 2D abelian topological orders.\\

\section{Lattice implementation}

We now implement the gauging of the 1-form symmetry on the $\R^N$ toric code, which serves as a lattice realization of $\R^N$ gauge theory~\cite{Magdalena2026}.

\subsection{The $\R^N$ toric code}

The Hilbert space $\mathcal H_{\R^N}$ consists of $N$ continuous-variable modes on each edge of the square lattice. Each mode has position and momentum operators $\hat{x}$ and $\hat{p}$ satisfying $[\hat{x},\hat{p}]=i$. We define displacement operators
\begin{equation}
    Z^t=e^{2\pi it\hat x},\qquad X^s=e^{-is\hat p},\qquad s,t\in\R,
\end{equation}
which obey the commutation relation
\begin{equation}
    Z^tX^s=e^{2\pi ist}X^sZ^t.
\end{equation}
Our normalization is chosen so that $Z$ shifts momentum by $2\pi$, mirroring the convention that $\phi=1$ denotes a $2\pi$ flux. We denote the displacement operators acting on the $n$th CV mode of edge $e$ by $X^s_{n,e}$ and $Z^t_{n,e}$, with $n=1,\ldots,N$.

The Hamiltonian of the $\R^N$ toric code is
\begin{equation} \label{eq:Rtoric}
    H_\mathrm{TC}=-\sum_{n,v}\int_{\R}d\phi\,A_{n,v}^\phi-\sum_{n,p}\int_{\R}dq\,B_{n,p}^q,
\end{equation}
where
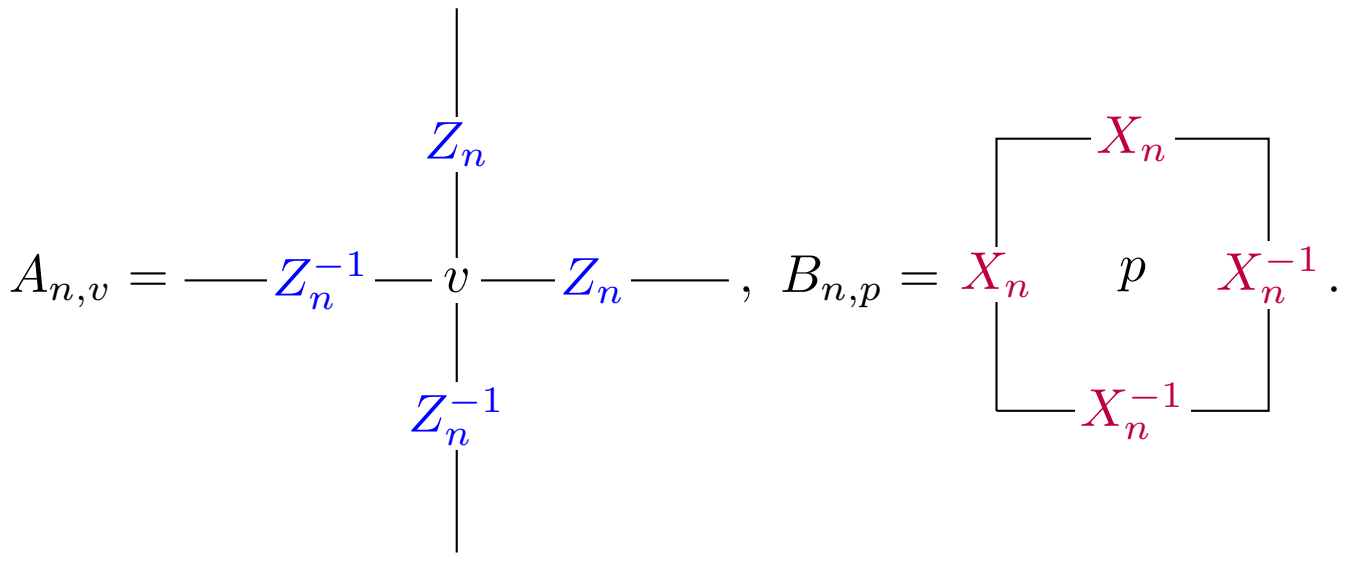
\begin{equation*}
    A_{n,v}=\begin{tikzpicture}[scale=1.8,baseline=-0.5ex]
        \draw (0,0)--(1,0);
        \draw (0,0)--(0,1);
        \draw (0,0)--(0,-1);
        \draw (0,0)--(-1,0);
        \node[draw=none,fill=white,inner sep=1pt] at (.5,0) {\color{blue}$Z_n$};
        \node[draw=none,fill=white,inner sep=1pt] at (0,.5) {\color{blue}$Z_n$};
        \node[draw=none,fill=white,inner sep=1pt] at (0,-.5) {\color{blue}$Z_n^{-1}$};
        \node[draw=none,fill=white,inner sep=1pt] at (-.5,0) {\color{blue}$Z_n^{-1}$};
        \node[draw=none,fill=white,inner sep=2pt] at (0,0) {$v$};
    \end{tikzpicture}\,,\,\,
    B_{n,p}=\begin{tikzpicture}[scale=1.8,baseline=5.2ex]
        \draw (0,0)--(1,0)--(1,1)--(0,1)--(0,0);
        \node[draw=none,fill=white,inner sep=1pt] at (.5,0) {\color{purple}$X_n^{-1}$};
        \node[draw=none,fill=white,inner sep=1pt] at (1,.5) {\color{purple}$X_n^{-1}$};
        \node[draw=none,fill=white,inner sep=1pt] at (.5,1) {\color{purple}$X_n$};
        \node[draw=none,fill=white,inner sep=1pt] at (0,.5) {\color{purple}$X_n$};
        \node[draw=none,fill=white,inner sep=1pt] at (.5,.5) {$p$};
    \end{tikzpicture}.
\end{equation*}
Violations of the vertex constraint are electric gauge charges, whereas violations of the plaquette constraint are magnetic gauge fluxes. In particular, an electric charge $q_n$ at vertex $v$ corresponds to eigenvalue $e^{2\pi i \alpha q_n}$ of $A_{n,v}^\alpha$ for all $\alpha$, and a magnetic flux $\phi_n$ through plaquette $p$ corresponds to eigenvalue $e^{2\pi i \beta \phi_n}$ of $B_{n,p}^\beta$ for all $\beta$.

The closed string operators associated with the dyons in $L$ form a non-on-site $\Z^{2N}$ $1$-form symmetry of the $\R^N$ toric code. Let $t_i\in\Z^{2N}$ denote the $i$th generator, corresponding to the dyon $c_i = M t_i \in L$. There is some freedom in the choice of the microscopic representative of the symmetry transformations associated to $t_i$. We consider the symmetry generated by operators
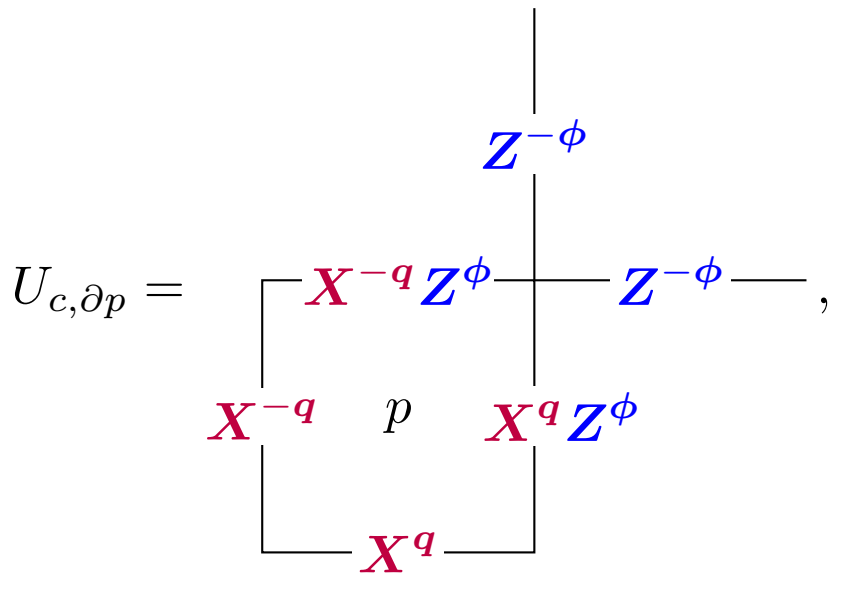
\begin{equation*}
    U_{c,\partial p}=\begin{tikzpicture}[scale=1.8,baseline=-1ex]
        \draw (0,0)--(-1,0)--(-1,-1)--(0,-1)--(0,0);
        \draw (0,0)--(0,1);
        \draw (0,0)--(1,0);
        \node[draw=none,fill=white,inner sep=0pt] at (-.5,0) {\color{purple}$\bs X^{-\bs q}$\color{blue}$\bs Z^{\bs\phi}$};
        \node[draw=none,fill=white,inner sep=1pt] at (.1,-.5) {\color{purple}$\bs X^{\bs  q}$\color{blue}$\bs Z^{\bs\phi}$};
        \node[draw=none,fill=white,inner sep=1pt] at (0,.5) {\color{blue}$\bs Z^{-\bs\phi}$};
        \node[draw=none,fill=white,inner sep=1pt] at (.5,0) {\color{blue}$\bs Z^{-\bs\phi}$};
        \node[draw=none,fill=white,inner sep=1pt] at (-1,-.5) {\color{purple}$\bs X^{-\bs q}$};
        \node[draw=none,fill=white,inner sep=1pt] at (-.5,-1) {\color{purple}$\bs X^{\bs q}$};
        \node[draw=none,fill=white,inner sep=1pt] at (-.5,-.5) {$p$};
    \end{tikzpicture}\,,
\end{equation*}
where
\begin{equation}
    \bs Z^{\bs\phi}=\prod_{n=1}^{N}Z_n^{\phi^n},\qquad\bs X^{\bs q}=\prod_{n=1}^{N}X_n^{q^n}.
\end{equation}

The 1-form symmetry generated by $U_{c_i,\partial p}$ is anomaly-free if $c_i$ is a boson, but anomalous if $c_i$ is a fermion~\cite{Gaiotto2015Generalized,Bhardwaj2017SpinTFT}. We denote the fermion parity $p_i$ of the dyon $c_i$ as 
\begin{equation}
    p_i = \langle c_i, c_i \rangle \pmod2.
\end{equation}
For each $i$ such that $c_i$ is fermionic, we introduce a physical fermionic mode at each vertex $v$ described by Majorana operators $\gamma_{i,v}$ and $\gamma_{i,v}'$. The Majorana modes form a Hilbert space $\mathcal{H}_f$. These ancillary modes trivialize the possible anomalies of the 1-form symmetry~\footnote{In principle, it would suffice to introduce a single mode on each vertex. We introduce a mode for each fermionic generator as a matter of convenience.}. We will gauge the augmented toric code Hamiltonian,
\begin{equation}
    H=H_\mathrm{TC}-\sum_{i,v}F_{i,v}^{p_i},\qquad F_{i,v}=i\gamma_{i,v}\gamma'_{i,v}.
\end{equation}
Here and below, a fermionic factor raised to $p_i=0$ denotes the identity.\\

\subsection{Gauging the 1-form symmetry}

Adapting the defect-based gauging approach of Ref.~\cite{Seifnashri2024Gauging}, our procedure consists of three steps:\\

1) Introduce a $\Z^{2N}$ 2-form gauge field living on the vertices of the lattice, forming a Hilbert space $\mathcal{H}_{\Z^{2N}}$ consisting of $2N$ rotor degrees of freedom on each vertex. The total Hilbert space is thus $\mathcal{H}=\mathcal{H}_{\R^{N}}\otimes\mathcal{H}_f\otimes\mathcal{H}_{\Z^{2N}}$. Each rotor has angular coordinate $\hat\theta$ and angular momentum $\hat{n}$ satisfying $[\hat\theta,\hat{n}]=i$. The eigenvalues of the $2N$ angular momentum operators on vertex $v$ encode the symmetry defect $g_v\in\Z^{2N}$ at $v$. The Hilbert space $\mathcal{H}_{\Z^{2N}}$ is thus spanned by defect configurations $\ket{\{g_v\}}$. We define displacement operators
\begin{equation}
    \mathsf{Z}^t=e^{2\pi it\hat n},\qquad\mathsf{X}^s=e^{is\hat\theta},\qquad s\in\Z,\,t\in\R,
\end{equation}
obeying the commutation relation
\begin{equation}
    \mathsf{Z}^t\mathsf{X}^s=e^{2\pi ist}\mathsf{X}^s\mathsf{Z}^t.
\end{equation}
In particular, $\mathsf{Z}^m=\mathbbm1$ for $m\in\Z$. We denote the displacement operators acting on the $i$th rotor on vertex $v$ by $\mathsf{Z}_{i,v}^t$ and $\mathsf{X}_{i,v}^s$, for $i=1,\ldots,2N$.\\

2) Impose Gauss law constraints $\mathcal{G}_{i,e}=1$ for each edge $e$, oriented from $u$ to $v$, and $i=1,\ldots,2N$, where the Gauss law operators satisfy the following properties:
\newline
\begin{enumerate}[leftmargin=*,label=\Roman*.,nosep]
    \item $\mathcal{G}_{i,e}=\mathsf{X}^{-1}_{i,u}\cdot\mathsf{X}_{i,v}\cdot V_{i,e}$ for some unitary operator $V_{i,e}$ diagonal in the $\ket{\{g_v\}}$ basis and supported near $e$.\\
    \item $\bra{\{g_v=0\}}\prod_{e\in\partial p}\mathcal{G}_{i,e}\ket{\{g_v=0\}}=U_{c_i,\partial p}$ in the bosonic case, and $U_{c_i,\partial p}\left(F_{i,\mathrm{SW}(p)}F_{i,\mathrm{NE}(p)}\right)^{p_i}$ in general.\\
    \item $\left[\mathcal{G}_{i,e},\mathcal{G}_{j,e'}\right]=0$.\\
\end{enumerate}
Taken together, properties I and II imply that $\mathcal{G}_{i,e}$ can be interpreted as a short string operator that hops a $t_i$ symmetry defect from $u$ to $v$. In the fermionic case, the additional Majorana factors in property II implement fermionic gauging by attaching a physical fermion to each fermionic symmetry defect. Property III ensures that the constraints $\mathcal{G}_{i,e}=1$ can be imposed simultaneously, either directly or energetically. To obtain a stabilizer code model, we will impose the constraints energetically.\\

3) Define the gauged Hamiltonian,
\begin{equation}
    \tilde H=\sum_{\{g_v\}}U_{\{g_v\}}HU_{\{g_v\}}^{-1}\otimes\ket{\{g_v\}}\bra{\{g_v\}}-\sum_{i,e}\left(\mathcal{G}_{i,e}+\mathcal{G}_{i,e}^\dagger\right),
\end{equation}
in terms of defect insertion operators $U_{\{g_v\}}$ defined for each defect configuration $\{g_v\}$. Specifically,
\begin{equation}
    U_{\{g_v\}}=\prod_{i,v}\,(U_{i,\gamma_v})^{g_v^i},
\end{equation}
in terms of elementary defect insertion operators
\begin{equation}
    U_{i,\gamma_v}=\bra{\{g_{v'}=\delta_{v,v'}t_i\}}\prod_{e\in \gamma_v}\mathcal{G}_{i,e}\ket{\{g_{v'}=0\}}.
\end{equation}
Here, $\gamma_v$ is an arbitrary choice of semi-infinite oriented path terminating at $v$. For concreteness, we take $\gamma_v$ to be a straight line from the west. Thus $U_{i,\gamma_v}$ creates a $t_i$ symmetry defect at vertex $v$. Note that these operators are well-defined due to commutativity of the Gauss law operators. By construction, $\tilde H$ commutes with the Gauss law operators.\\

We carry out this procedure explicitly by choosing the Gauss law operators 
\begin{equation}
    \mathcal{G}_{i,e}=\begin{tikzpicture}[scale=2,baseline=5.9ex]
        \draw (0,0)--(0,1)--(1,1);
        \node[draw=none,fill=white,inner sep=1pt] at (0,.5) {\color{purple}$\bs{X}_e^{\bs q_i}$};
        \node[draw=none,fill=white,inner sep=1pt] at (.5,1) {\color{blue}$\bs{Z}^{-\bs\phi_i}$};
        \node[draw=none,fill=white,inner sep=1pt] at (0,0) {$Q_i^{-1}$};
        \node[draw=none,fill=white,inner sep=1pt] at (0,1) {$Q_i'$};
    \end{tikzpicture},\,\quad
    \begin{tikzpicture}[scale=2,baseline=5.9ex]
        \draw (0,0)--(1,0)--(1,1);
        \node[draw=none,fill=white,inner sep=1pt] at (1,.5) {\color{blue}$\bs{Z}^{\bs\phi_i}$};
        \node[draw=none,fill=white,inner sep=1pt] at (0,0) {$Q_i'^{-1}$};
        \node[draw=none,fill=white,inner sep=2pt] at (1,0) {$Q_i$};
        \node[draw=none,fill=white,inner sep=1pt] at (.55,0){\color{purple}$\bs{X}_e^{\bs q_i}$};
    \end{tikzpicture},
\end{equation}
where \footnote{For fermions, the ordering of the fermionic operators must be specified. We pick the convention that the tail endpoint factor is ordered to the left of the head endpoint factor in each diagram.}
\begin{align}
    Q_i&=\mathsf{X}_i\gamma^{p_i}_i\prod_{j<i}\mathsf{Z}_j^{-\bs\phi_j\cdot\bs q_i},\\ Q_i'&=\mathsf{X}_i\gamma'^{p_i}_i\prod_{j>i}\mathsf{Z}_j^{\bs\phi_j\cdot\bs q_i}.
\end{align}
It is straightforward to verify that these operators satisfy properties I--III. The Majorana dressing is closely related to exact two-dimensional bosonization~\cite{Chen2018Bosonization}; products of Majorana hopping operators around a plaquette yield fermion parities at two opposite corners, explaining the parity factors in property II.

The resulting defect insertion operator has the form
\begin{equation}
    U_{i,\gamma_v}=(i\gamma'_{i,v})^{p_i}\left(\prod_{w\in\gamma_v}F_{i,w}^{p_i}\right)\left(\prod_{e\in\gamma_v}\bs{X}_e^{\bs q_i}\bs{Z}_{\mathrm{NE}(e)}^{\bs\phi_i}\right),
\end{equation}
where $\mathrm{NE}(e)$ denotes the edge situated to the northeast of $e$. This operator commutes with $A_{n,v'}$ unless $v=v'$, with $B_{n,p}$ unless $v$ is the southwest corner of $p$, and with $F_{i,v'}$ unless $v=v'$ and $p_i=1$. Imposing the Gauss law constraints $\mathcal{G}_{i,e}=1$ energetically, the gauged Hamiltonian is therefore
\begin{equation}
\begin{split}
    \tilde H=-\sum_{n,v}\int_{\R}d\phi\,\tilde{A}_{n,v}^\phi-\sum_{n,p}\int_{\R}dq\,\tilde{B}_{n,p}^q\\-\sum_{i,v}\tilde F_{i,v}^{p_i}-\sum_{i,e}\left(\mathcal{G}_{i,e}+\mathcal{G}_{i,e}^\dagger\right),
\end{split}
\label{eq:Htilde}
\end{equation}
where \footnote{The powers $\tilde A^\phi$ and $\tilde B^q$ are defined by multiplying every Pauli exponent in the expressions for $\tilde A$ and $\tilde B$ by $\phi$ and $q$, respectively, before simplifying any factors that equal the identity. For instance, a rotor factor $\mathsf Z^m$ with $m\in\Z$ contributes $\mathsf Z^{m\phi}$ to $\tilde A^\phi$.}
\begin{equation*}
    \tilde A_{n,v}=\begin{tikzpicture}[scale=1.8,baseline=-0.5ex]
        \draw (0,0)--(1,0);
        \draw (0,0)--(0,1);
        \draw (0,0)--(0,-1);
        \draw (0,0)--(-1,0);
        \node[draw=none,fill=white,inner sep=1pt] at (.6,0) {\color{blue}$Z_n$};
        \node[draw=none,fill=white,inner sep=1pt] at (0,.6) {\color{blue}$Z_n$};
        \node[draw=none,fill=white,inner sep=1pt] at (0,-.6) {\color{blue}$Z_n^{-1}$};
        \node[draw=none,fill=white,inner sep=1pt] at (-.6,0) {\color{blue}$Z_n^{-1}$};
        \node[draw=none,fill=white,inner sep=1pt] at (.02,0) {$\bs{\mathsf{Z}}_v^{\vec q^{\,n}}$};
    \end{tikzpicture}\,,\,\,
    \tilde B_{n,p}=\begin{tikzpicture}[scale=1.8,baseline=5.2ex]
        \draw (0,0)--(1,0)--(1,1)--(0,1)--(0,0);
        \node[draw=none,fill=white,inner sep=1pt] at (.5,0) {\color{purple}$X_n^{-1}$};
        \node[draw=none,fill=white,inner sep=1pt] at (1,.5) {\color{purple}$X_n^{-1}$};
        \node[draw=none,fill=white,inner sep=1pt] at (.5,1) {\color{purple}$X_n$};
        \node[draw=none,fill=white,inner sep=1pt] at (0,.5) {\color{purple}$X_n$};
        \node[draw=none,fill=white,inner sep=1pt] at (.02,0) {$\bs{\mathsf{Z}}^{\vec\phi^n}$};
        \node[draw=none,fill=white,inner sep=1pt] at (.5,.5) {$p$};
    \end{tikzpicture},
\end{equation*}
and
\begin{equation}
    \tilde F_{i,v}=i\gamma_{i,v}\gamma'_{i,v}\mathsf{Z}_{i,v}^{\frac{1}{2}}.
\end{equation}
Here we have introduced the notation
\begin{equation}
    \bs{\mathsf{Z}}^{\vec\phi^n}=\prod_{i=1}^{2N}\mathsf{Z}_i^{\phi_i^n},\qquad
    \bs{\mathsf{Z}}^{\vec{q}^{\,n}}=\prod_{i=1}^{2N}\mathsf{Z}_i^{q_i^n}.
\end{equation}

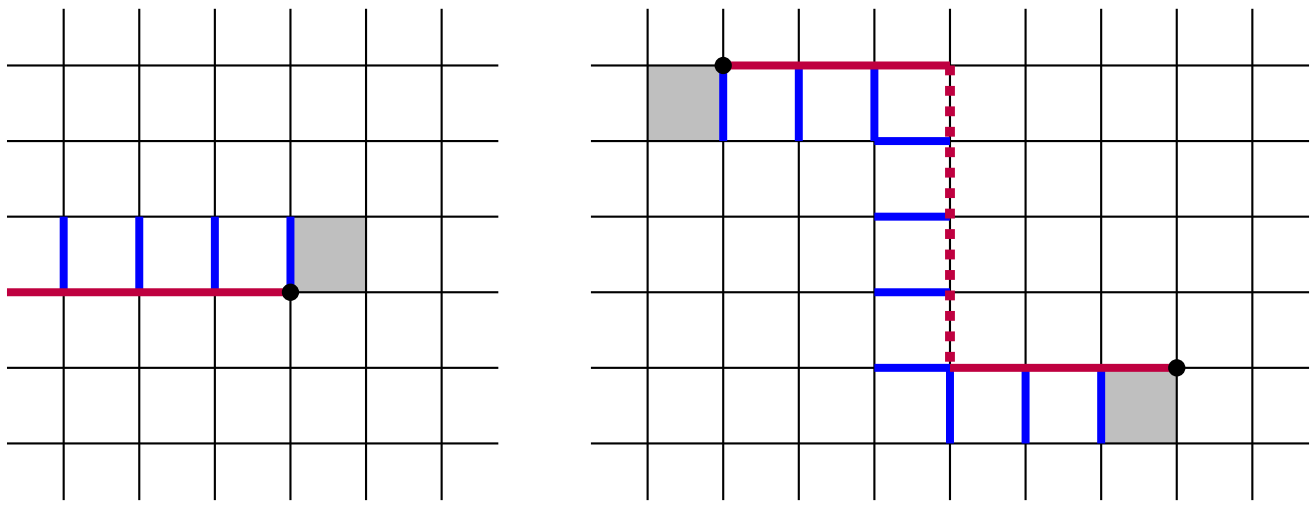
\begin{figure}[!t]
    \centering
    \begin{tikzpicture}[scale=.5]
        \fill[color=lightgray] (4,3) rectangle (5,4);
        \foreach \x in {1,...,6} {
            \draw[thin] (\x,.25)--(\x,6.75);
        }
        \foreach \y in {1,...,6} {
            \draw[thin] (.25,\y)--(6.75,\y);
        }
        \foreach \x in {1,...,4} {
            \draw[color=blue,line width=1.5] (\x,3)--(\x,4);
        }
        \draw[color=purple,line width=1.5](.25,3)--(4,3);
        \fill (4,3) circle (3.25pt);
    \end{tikzpicture}\hspace{14pt}
    \begin{tikzpicture}[scale=.5]
        \fill[color=lightgray] (0,5) rectangle (1,6);
        \fill[color=lightgray] (7,1) rectangle (6,2);
        \foreach \x in {0,...,8} {
            \draw[thin] (\x,.25)--(\x,6.75);
        }
        \foreach \y in {1,...,6} {
            \draw[thin] (-.75,\y)--(8.75,\y);
        }
        \foreach \x in {1,...,3} {
            \draw[color=blue,line width=1.5] (\x,5)--(\x,6);
        }
        \foreach \y in {2,...,5} {
            \draw[color=blue,line width=1.5] (3,\y)--(4,\y);
        }
        \foreach \x in {4,...,6} {
            \draw[color=blue,line width=1.5] (\x,1)--(\x,2);
        }
        \draw[color=purple,line width=1.5](1,6)--(4,6);
        \draw[color=purple,line width=1.85,dotted](4,6)--(4,2);
        \draw[color=purple,line width=1.5](4,2)--(7,2);
        \fill (1,6) circle (3.25pt);
        \fill (7,2) circle (3.25pt);
    \end{tikzpicture}
    \caption{(a) The defect insertion operator $U_{i,\gamma_v}$, which acts as $\bs Z^{\bs\phi_i}$ on solid blue edges and $\bs X^{\bs q_i}$ on solid red    edges. The black dot represents vertex $v$. The Majorana factors are suppressed. (b) An open string operator $W_{a,\gamma}$ for the deconfined dyon $a=(\bs\psi,\bs r)\in L^\vee$, acting as $\bs Z^{\bs\psi}$ on solid blue edges and $\bs X^{\pm\bs r}$ on solid/dashed red edges. $W_{a,\gamma}$ excites the plaquette and vertex term at each endpoint.}
    \label{fig:string}
\end{figure}

\subsection{Topological order of the gauged model}

We now examine how the dyon condensation rules arise microscopically in the gauged model. String operators for generic dyons of the original $\mathbb{R}^N$ toric code fail to commute with the Gauss law operators along their length, reflecting their confinement. The dyons in $L^\vee$, however, admit string operators that commute with the Hamiltonian away from their endpoints.

To construct these strings, consider the product
\begin{equation}
    W_{a,\partial p}=\prod_n\tilde A_{n,v}^{\psi^n}\tilde B_{n,p}^{r^n},
\end{equation}
where $a=(\boldsymbol{\psi},\boldsymbol{r})$ and $v$ is the southwest corner of $p$. Its action on the $j$th rotor at $v$ is $\mathsf{Z}^{\langle a,c_j\rangle}$. For $a\in L^\vee$, this action is trivial, leaving the closed string
\begin{equation}
    W_{a,\partial p}=\begin{tikzpicture}[scale=1.8,baseline=-0.5ex]
        \draw (0,0)--(1,0)--(1,1)--(0,1)--(0,0);
        \draw (0,0)--(0,-1);
        \draw (0,0)--(-1,0);
        \node[draw=none,fill=white,inner sep=1pt] at (.5,0) {\color{blue}$\bs Z^{\bs\psi}$\color{purple}$\bs X^{-\bs r}$};
        \node[draw=none,fill=white,inner sep=1pt] at (-.05,.5) {\color{blue}$\bs Z^{\bs\psi}$\color{purple}$\bs X^{\bs r}$};
        \node[draw=none,fill=white,inner sep=1pt] at (0,-.5) {\color{blue}$\bs Z^{-\bs\psi}$};
        \node[draw=none,fill=white,inner sep=1pt] at (-.5,0) {\color{blue}$\bs Z^{-\bs\psi}$};
        \node[draw=none,fill=white,inner sep=1pt] at (.5,1) {\color{purple}$\bs X^{\bs r}$};
        \node[draw=none,fill=white,inner sep=1pt] at (1,.5) {\color{purple}$\bs X^{-\bs r}$};
        \node[draw=none,fill=white,inner sep=1pt] at (.5,.5) {$p$};
    \end{tikzpicture}.
\end{equation}
Multiplying these elementary loops over a region cancels their action on the interior edges, giving a string supported near the boundary. For $a\notin L^\vee$, the same construction instead leaves a nontrivial action on rotors throughout the interior.

The corresponding open string along an oriented path $\gamma$ is
\begin{equation} 
    W_{a,\gamma} = \left(\prod_{e\in\gamma}\boldsymbol{Z}_{\mathrm{SW}(e)}^{\eta(e)\epsilon(e)\boldsymbol{\psi}}\right)\left(\prod_{e\in\gamma}\boldsymbol{X}_e^{-\epsilon(e)\boldsymbol{r}}\right).
    \label{eq:deconfined-string}
\end{equation}
Here $\mathrm{SW}(e)$ is the edge situated to the southwest of $e$, $\epsilon(e)=+1$ for eastward or northward steps and $-1$ for westward or southward steps, and $\eta(e)=-1$ for horizontal edges and $+1$ for vertical edges. An example is shown in Fig.~\ref{fig:string}(b). Each elementary segment commutes with every Gauss law operator when $\langle a,c_i\rangle\in\mathbb{Z}$ for all $i$. The string therefore creates excitations only at its endpoints: a dyon $-a$ at the initial endpoint and $a$ at the final endpoint, with the flux at each endpoint residing on the plaquette southwest of its charge. These are the deconfined anyons predicted by the dyon condensation picture.

A pair of deconfined anyons $a,a'\in L^\vee$ belong to the same superselection sector, modulo local excitations including physical fermions, whenever $a'-a\in L$. To see why, consider the local operator
\begin{equation}
    D_{i,v}=\begin{tikzpicture}[scale=1.8,baseline=5ex]
        \draw[dotted] (0,0)--(1,0)--(1,1);
        \draw (0,0)--(0,1)--(1,1);
        \node[fill=white,inner sep=1pt] at (.5,1) {\color{purple}$\boldsymbol{X}^{-\boldsymbol{q}_i}$};
        \node[fill=white,inner sep=2pt] at (0,0) {$Q_i$};
        \node[fill=white,inner sep=1pt] at (0,.5) {\color{purple}$\boldsymbol{X}^{-\boldsymbol{q}_i}$};
        \node[fill=white,inner sep=1pt] at (1,1) {$v$};
        \node[fill=white,inner sep=1pt] at (.5,.5) {$p$};
    \end{tikzpicture}.
    \label{eq:local-condensate}
\end{equation}
The rotor shift in $Q_i$ creates charge $\boldsymbol{q}_i$ at the southwest corner of $p$ and flux $2\pi\boldsymbol{\phi}_i$ through $p$, while the two CV shifts move the charge to $v$. Moreover, $D_{i,v}$ commutes with every Gauss law operator and every $\tilde F_{j,v'}$. It therefore creates precisely the endpoint excitation of a $c_i$ string at $v$: a flux through the plaquette southwest of a charge. Thus, applying $D_{i,v}$ at a string endpoint transforms $a$ into $a+c_i$, identifying dyon labels modulo $L$. We conclude that the anyons of the gauged model have fusion group $\mathcal{A}=L^\vee/L$, with braiding statistics inherited from the original dyons.

We compute the ground state degeneracy of the purely bosonic model on a torus in Appendix~\ref{app:gsd}. We find that $\mathrm{GSD}=|\mathcal{A}|$, with logical operators given by noncontractible strings associated with dyons in $L^\vee$. Together with the fusion rules and braiding statistics, this calculation confirms that the model realizes the topological order described by the input $K$-matrix.\\

\section{Discussion}

We have constructed CV stabilizer code realizations of all achiral 2D abelian topological orders. Our construction implements dyon condensation by gauging a non-on-site 1-form symmetry; a general formalism for gauging anomaly-free 1-form symmetries will be discussed in future work~\cite{QiShirleyForthcoming}. Conversely, we believe that chiral topological orders cannot be realized by CV stabilizer codes purely in two spatial dimensions. On the other hand, we do expect it is possible to realize any abelian topological order at the surface of a 3D CV stabilizer code with no deconfined bulk excitations. We conjecture that such 3D stabilizer codes can be disentangled by CV Clifford quantum cellular automata (QCAs), which are nontrivial in the Clifford category precisely when the surface topological order is chiral. Related connections between surface topological order and QCAs have been established for qudit systems~\cite{HaahFidkowskiHastings2023QCA,Haah2021CliffordQCA,Shirley2022QCA,Yang2026CliffordQCA}.

The Hamiltonians in this paper should be understood as ideal CV stabilizer models: exact codewords involve non-normalizable states, as in multimode GKP codes~\cite{Gottesman2001Oscillator}. 
An additional ingredient here is the presence of continuous stabilizers, whose integrals formally impose exact constraints. We expect that these constraints can be softened into finite-strength energy penalties by cutting off the domain of integration. As in Ref.~\cite{Magdalena2026}, we expect that adding suitable quadratic terms in the CV variables and rotor angular momenta will produce normalizable states with a finite energy gap while preserving the desired topological order.\\

\textbf{Acknowledgments}---We are grateful to Clay Cordova, Michael Levin, and Valerie Wu for insightful discussions. The authors used various versions of ChatGPT and Claude to assist with drafting and revising exposition, as well as exploring, developing, and checking mathematical arguments. The work of MQ is supported in part by the Simons Collaboration on Ultra-Quantum Matter, which is a grant from the Simons Foundation (Grant No. 651442, D.T.S.). The work of WS is supported by U.S. Department of Energy Grant 5-29073. The authors of this paper were ordered alphabetically.

\appendix

\onecolumngrid

\section{Stabilizer models of \texorpdfstring{$2\times 2$ even $K$-matrices}{2 by 2 even K-matrices}}

In this appendix, we specialize the construction of the main text to the $N=1$ bosonic case, then describe two alternative CV stabilizer code constructions of bosonic topological orders described by a $2\times 2$ even $K$-matrix. Let $c_i=(\phi_i,q_i)$, $i=1,2$, generate the condensate lattice $L\subset\R^2$, with $K_{ij}=\phi_iq_j+q_i\phi_j$. The Hilbert space $\mathcal{H}=\mathcal{H}_{\R}\otimes\mathcal{H}_{\Z^2}$ is composed of one CV mode per edge and two gauge rotors per vertex.

In this case, the gauged Hamiltonian \eqref{eq:Htilde} is
\begin{equation}
    \tilde H=-\sum_v\int_{\R}d\alpha\,\tilde A_v^\alpha-\sum_p\int_{\R}d\beta\,\tilde B_p^\beta-\sum_{i=1,2}\sum_e\left(\mathcal{G}_{i,e}+\mathcal{G}_{i,e}^\dagger\right).
    \label{eq:appA-gauged}
\end{equation}
The gauged star and plaquette operators are
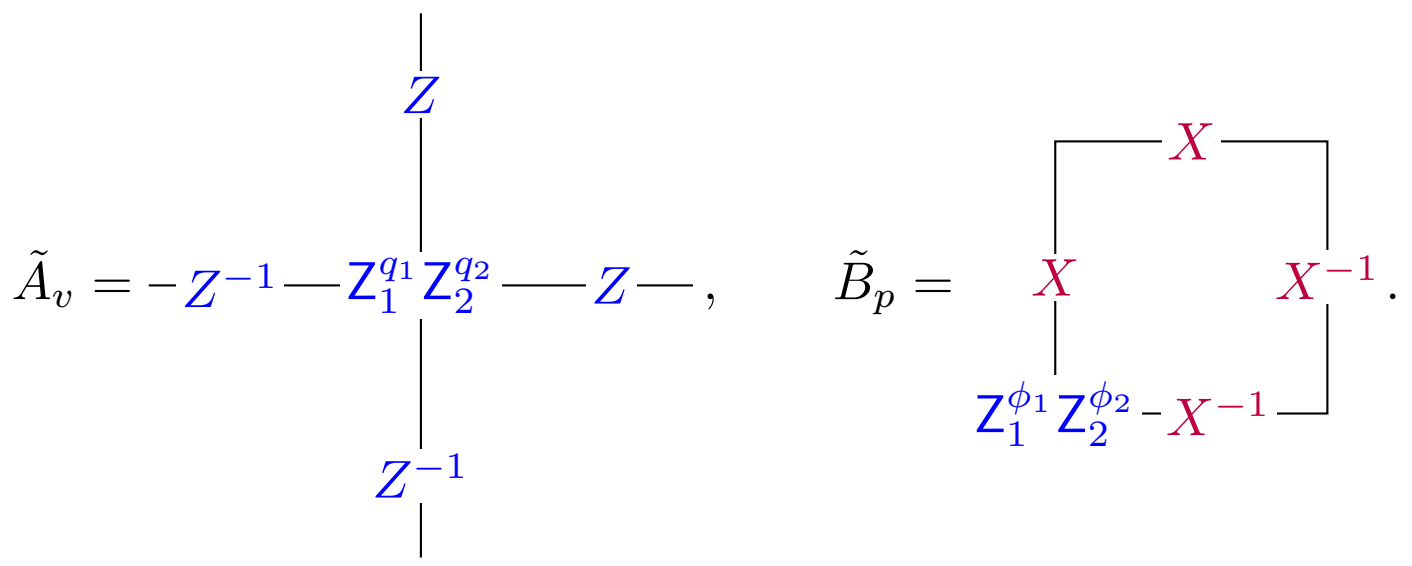
\begin{equation}
    \tilde A_v=\begin{tikzpicture}[scale=1.8,baseline=-.6ex]
        \draw[color=black] (0,0)--(1,0);
        \draw[color=black] (0,0)--(0,1);
        \draw[color=black] (0,0)--(0,-1);
        \draw[color=black] (0,0)--(-1,0);
        \node[draw=none,fill=white,inner sep=1pt] at (.7,0) {\color{blue}$Z$};
        \node[draw=none,fill=white,inner sep=1pt] at (0,.7) {\color{blue}$Z$};
        \node[draw=none,fill=white,inner sep=1pt] at (0,-.7) {\color{blue}$Z^{-1}$};
        \node[draw=none,fill=white,inner sep=1pt] at (-.7,0) {\color{blue}$Z^{-1}$};
        \node[draw=none,fill=white,inner sep=1pt] at (0,0) {\color{blue}$\mathsf{Z}_1^{q_1}\mathsf{Z}_2^{q_2}$};
    \end{tikzpicture}\,,\qquad
    \tilde B_p=\begin{tikzpicture}[scale=1.8,baseline=5ex]
        \draw[color=black] (0,0)--(1,0)--(1,1)--(0,1)--(0,0);
        \node[draw=none,fill=white,inner sep=1pt] at (.6,0) {\color{purple}$X^{-1}$};
        \node[draw=none,fill=white,inner sep=1pt] at (1,.5) {\color{purple}$X^{-1}$};
        \node[draw=none,fill=white,inner sep=1pt] at (.5,1) {\color{purple}$X$};
        \node[draw=none,fill=white,inner sep=1pt] at (0,.5) {\color{purple}$X$};
        \node[draw=none,fill=white,inner sep=1pt] at (0,0) {\color{blue}$\mathsf{Z}_1^{\phi_1}\mathsf{Z}_2^{\phi_2}$};
    \end{tikzpicture}.
\end{equation}
The Gauss law operators are
\begin{equation}
    \mathcal{G}_{1,e}=
    \begin{tikzpicture}[scale=1.8,baseline=5.6ex]
        \draw[color=black] (0,0)--(0,1)--(1,1);
        \node[draw=none,fill=white,inner sep=1pt] at (0,.5) {\color{purple}$X_e^{q_1}$};
        \node[draw=none,fill=white,inner sep=1pt] at (.5,1) {\color{blue}$Z^{-\phi_1}$};
        \node[draw=none,fill=white,inner sep=1pt] at (0,1) {$Q_1'$};
        \node[draw=none,fill=white,inner sep=1pt] at (0,0) {$Q_1^{-1}$};
    \end{tikzpicture},\quad
    \begin{tikzpicture}[scale=1.8,baseline=5.6ex]
        \draw[color=black] (0,0)--(1,0)--(1,1);
        \node[draw=none,fill=white,inner sep=0pt] at (.5,0) {\color{purple}$X_e^{q_1}$};
        \node[draw=none,fill=white,inner sep=1pt] at (1,.5) {\color{blue}$Z^{\phi_1}$};
        \node[draw=none,fill=white,inner sep=0pt] at (0,0) {$Q_1'^{-1}$};
        \node[draw=none,fill=white,inner sep=1pt] at (1,0) {$Q_1$};
    \end{tikzpicture},\qquad
    \mathcal{G}_{2,e}=
    \begin{tikzpicture}[scale=1.8,baseline=5.6ex]
        \draw[color=black] (0,0)--(0,1)--(1,1);
        \node[draw=none,fill=white,inner sep=1pt] at (0,.5) {\color{purple}$X_e^{q_2}$};
        \node[draw=none,fill=white,inner sep=1pt] at (.5,1) {\color{blue}$Z^{-\phi_2}$};
        \node[draw=none,fill=white,inner sep=1pt] at (0,1) {$Q'_2$};
        \node[draw=none,fill=white,inner sep=1pt] at (0,0) {$Q_2^{-1}$};
    \end{tikzpicture},\quad
    \begin{tikzpicture}[scale=1.8,baseline=5.6ex]
        \draw[color=black] (0,0)--(1,0)--(1,1);
        \node[draw=none,fill=white,inner sep=0pt] at (.5,0) {\color{purple}$X_e^{q_2}$};
        \node[draw=none,fill=white,inner sep=1pt] at (1,.5) {\color{blue}$Z^{\phi_2}$};
        \node[draw=none,fill=white,inner sep=0pt] at (0,0) {$Q'^{-1}_2$};
        \node[draw=none,fill=white,inner sep=1pt] at (1,0) {$Q_2$};
    \end{tikzpicture},
\end{equation}
where
\begin{equation}
    Q_1=\mathsf{X}_1,\qquad Q'_1=\mathsf{X}_1\mathsf{Z}_2^{\phi_2q_1},\qquad Q_2=\mathsf{X}_2\mathsf{Z}_1^{-\phi_1q_2},\qquad Q'_2=\mathsf{X}_2.
\end{equation}

The first alternative construction uses a simplified choice of Gauss law operators on $\mathcal{H}=\mathcal{H}_{\R}\otimes\mathcal{H}_{\Z^2}$. In particular, we define
\begin{equation}
    \mathcal{G}'_{1,e}=
    \begin{tikzpicture}[scale=1.8,baseline=5.6ex]
        \draw (0,0)--(0,1)--(1,1);
        \node[draw=none,fill=white,inner sep=1pt] at (0,.5) {\color{purple}$X_e^{q_1}$};
        \node[draw=none,fill=white,inner sep=1pt] at (.5,1) {\color{blue}$Z^{-\phi_1}$};
        \node[draw=none,fill=white,inner sep=1pt] at (0,1) {\color{purple}$\mathsf{X}_1$};
        \node[draw=none,fill=white,inner sep=1pt] at (0,0) {\color{purple}$\mathsf{X}_1^{-1}$};
    \end{tikzpicture},\quad
    \begin{tikzpicture}[scale=1.8,baseline=5.6ex]
        \draw (0,0)--(1,0)--(1,1);
        \node[draw=none,fill=white,inner sep=0pt] at (.5,0) {\color{purple}$X_e^{q_1}$};
        \node[draw=none,fill=white,inner sep=1pt] at (1,.5) {\color{blue}$Z^{\phi_1}$};
        \node[draw=none,fill=white,inner sep=0pt] at (0,0) {\color{purple}$\mathsf{X}_1^{-1}$};
        \node[draw=none,fill=white,inner sep=1pt] at (1,0) {\color{purple}$\mathsf{X}_1$};
    \end{tikzpicture},\qquad
    \mathcal{G}'_{2,e}=
    \begin{tikzpicture}[scale=1.8,baseline=5.6ex]
        \draw (0,0)--(1,0)--(1,1);
        \node[draw=none,fill=white,inner sep=1pt] at (.45,0) {\color{blue}$Z^{-\phi_2}$};
        \node[draw=none,fill=white,inner sep=1pt] at (1,.5) {\color{purple}$X_e^{q_2}$};
        \node[draw=none,fill=white,inner sep=1pt] at (1,1) {\color{purple}$\mathsf{X}_2$};
        \node[draw=none,fill=white,inner sep=1pt] at (1,0) {\color{purple}$\mathsf{X}_2^{-1}$};
    \end{tikzpicture},\quad
    \begin{tikzpicture}[scale=1.8,baseline=5.6ex]
        \draw (0,0)--(0,1)--(1,1);
        \node[draw=none,fill=white,inner sep=1pt] at (0,.5) {\color{blue}$Z^{\phi_2}$};
        \node[draw=none,fill=white,inner sep=1pt] at (.5,1) {\color{purple}$X_e^{q_2}$};
        \node[draw=none,fill=white,inner sep=1pt] at (0,1) {\color{purple}$\mathsf{X}_2^{-1}$};
        \node[draw=none,fill=white,inner sep=1pt] at (1,1) {\color{purple}$\mathsf{X}_2$};
    \end{tikzpicture}.
\end{equation}
With this choice, the toric code operators $A_v$ and $B_p$ have gauged counterparts
\begin{equation}
    \tilde A'_v=\begin{tikzpicture}[scale=1.8,baseline=-0.5ex]
        \draw (0,0)--(1,0);
        \draw (0,0)--(0,1);
        \draw (0,0)--(0,-1);
        \draw (0,0)--(-1,0);
        \node[draw=none,fill=white,inner sep=1pt] at (.65,0) {\color{blue}$Z$};
        \node[draw=none,fill=white,inner sep=1pt] at (0,.6) {\color{blue}$Z$};
        \node[draw=none,fill=white,inner sep=1pt] at (0,-.6) {\color{blue}$Z^{-1}$};
        \node[draw=none,fill=white,inner sep=1pt] at (-.65,0) {\color{blue}$Z^{-1}$};
        \node[draw=none,fill=white,inner sep=1pt] at (0,0) {\color{blue}$\mathsf{Z}_1^{q_1}\mathsf{Z}_2^{q_2}$};
    \end{tikzpicture}\,,\quad
    \tilde B'_p=\begin{tikzpicture}[scale=1.8,baseline=5ex]
        \draw (0,0)--(1,0)--(1,1)--(0,1)--(0,0);
        \node[draw=none,fill=white,inner sep=1pt] at (.5,0) {\color{purple}$X^{-1}$};
        \node[draw=none,fill=white,inner sep=1pt] at (1,.5) {\color{purple}$X^{-1}$};
        \node[draw=none,fill=white,inner sep=1pt] at (.5,1) {\color{purple}$X$};
        \node[draw=none,fill=white,inner sep=1pt] at (0,.5) {\color{purple}$X$};
        \node[draw=none,fill=white,inner sep=1pt] at (0,0) {\color{blue}$\mathsf{Z}_1^{\phi_1}$};
        \node[draw=none,fill=white,inner sep=1pt] at (1,1) {\color{blue}$\mathsf{Z}_2^{\phi_2}$};
        \node[draw=none,fill=white,inner sep=1pt] at (.5,.5) {$p$};
    \end{tikzpicture}.
\end{equation}
The gauged Hamiltonian, with energetically imposed Gauss law constraints, is thus
\begin{equation}
    \tilde H'=-\sum_v\int_{\R}d\alpha\,\tilde A_v'^{\alpha}-\sum_p\int_{\R}d\beta\,\tilde B_p'^{\beta}-\sum_{i=1,2}\sum_e\left(\mathcal{G}'_{i,e}+\mathcal{G}'^\dagger_{i,e}\right).
\end{equation}

The second alternative construction implements condensation of $c_1$ and $c_2$ directly in $\mathcal{H}_{\R}$, without introducing gauge rotors. Choose a basis $a_i=(\psi_i,r_i)$, $i=1,2$, of the dual lattice $L^\vee$. The Hamiltonian is composed of elementary hopping operators $C_{i,e}$ for the condensed dyons and elementary loop operators $W_{a_i,\partial p}$ for the surviving dyons:
\begin{equation}
    H_c=-\sum_{i=1,2}\sum_p\left(W_{a_i,\partial p}+W^\dagger_{a_i,\partial p}\right)-\sum_{i=1,2}\sum_e\left(C_{i,e}+C^\dagger_{i,e}\right),
\end{equation}
where
\begin{equation}
    W_{a_i,\partial p}=
    \begin{tikzpicture}[scale=1.8,baseline=-.6ex]
        \draw (0,0)--(1,0)--(1,1)--(0,1)--(0,0);
        \draw (0,0)--(0,-1);
        \draw (0,0)--(-1,0);
        \node[draw=none,fill=white,inner sep=1pt] at (.5,0) {\color{purple}$X^{-r_i}$\color{blue}$Z^{\psi_i}$};
        \node[draw=none,fill=white,inner sep=1pt] at (-.05,.5) {\color{purple}$X^{r_i}$\color{blue}$Z^{\psi_i}$};
        \node[draw=none,fill=white,inner sep=1pt] at (0,-.5) {\color{blue}$Z^{-\psi_i}$};
        \node[draw=none,fill=white,inner sep=1pt] at (-.5,0) {\color{blue}$Z^{-\psi_i}$};
        \node[draw=none,fill=white,inner sep=1pt] at (.5,1) {\color{purple}$X^{r_i}$};
        \node[draw=none,fill=white,inner sep=1pt] at (1,.5) {\color{purple}$X^{-r_i}$};
        \node[draw=none,fill=white,inner sep=1pt] at (.5,.5) {$p$};
    \end{tikzpicture}
\end{equation}
and
\begin{equation}
    C_{1,e}=
    \begin{tikzpicture}[scale=1.8,baseline=-.6ex]
        \draw (0,0)--(0,1)--(1,1);
        \draw (0,0)--(1,0);
        \draw (0,0)--(-1,0);
        \draw (0,0)--(0,-1);
        \node[draw=none,fill=white,inner sep=1pt] at (0,.5) {\color{purple}$X_e^{q_1}$\color{blue}$Z^{\phi_1}$};
        \node[draw=none,fill=white,inner sep=1pt] at (.5,1) {\color{blue}$Z^{-\phi_1}$};
        \node[draw=none,fill=white,inner sep=1pt] at (.5,0) {\color{blue}$Z^{\phi_1}$};
        \node[draw=none,fill=white,inner sep=1pt] at (-.5,0) {\color{blue}$Z^{-\phi_1}$};
        \node[draw=none,fill=white,inner sep=1pt] at (0,-.5) {\color{blue}$Z^{-\phi_1}$};
    \end{tikzpicture}\,,\quad
    \begin{tikzpicture}[scale=1.8,baseline=-.6ex]
        \draw (0,0)--(1,0);
        \draw (0,0)--(-1,0);
        \draw (0,0)--(0,-1);
        \node[draw=none,fill=white,inner sep=1pt] at (0,-.5) {\color{blue}$Z^{\phi_1}$};
        \node[draw=none,fill=white,inner sep=1pt] at (-.5,0) {\color{purple}$X_e^{q_1}$\color{blue}$Z^{\phi_1}$};
        \node[draw=none,fill=white,inner sep=0pt] at (.5,0) {\color{blue}$Z^{-\phi_1}$};
    \end{tikzpicture}\,,\qquad
    C_{2,e}=
    \begin{tikzpicture}[scale=1.8,baseline=-.6ex]
        \draw (0,0)--(0,1);
        \draw (0,0)--(0,-1);
        \draw (0,0)--(-1,0);
        \node[draw=none,fill=white,inner sep=1pt] at (-.5,0) {\color{blue}$Z^{-\phi_2}$};
        \node[draw=none,fill=white,inner sep=1pt] at (0,-.5) {\color{purple}$X_e^{q_2}$\color{blue}$Z^{-\phi_2}$};
        \node[draw=none,fill=white,inner sep=1pt] at (0,.5) {\color{blue}$Z^{\phi_2}$};
    \end{tikzpicture},\quad
    \begin{tikzpicture}[scale=1.8,baseline=-.6ex]
        \draw (0,0)--(1,0)--(1,1);
        \draw (0,0)--(0,1);
        \draw (0,0)--(0,-1);
        \draw (0,0)--(-1,0);
        \node[draw=none,fill=white,inner sep=1pt] at (.5,0) {\color{purple}$X_e^{q_2}$\color{blue}$Z^{-\phi_2}$};
        \node[draw=none,fill=white,inner sep=1pt] at (1,.5) {\color{blue}$Z^{\phi_2}$};
        \node[draw=none,fill=white,inner sep=1pt] at (0,.5) {\color{blue}$Z^{-\phi_2}$};
        \node[draw=none,fill=white,inner sep=1pt] at (0,-.5) {\color{blue}$Z^{\phi_2}$};
        \node[draw=none,fill=white,inner sep=1pt] at (-.5,0) {\color{blue}$Z^{\phi_2}$};
    \end{tikzpicture}.
\end{equation}

\section{Ground state degeneracy on the torus}
\label{app:gsd}

In this appendix, we compute the logical operators and ground state degeneracy of the gauged model~\eqref{eq:Htilde} in the case of bosonic $L$, corresponding to a topological order described by an even signature-zero $K$ matrix. Logical operators are closed string operators for deconfined dyons in $L^\vee$ acting on noncontractible cycles, and closed (contractible and noncontractible) string operators for condensed dyons in $L$ are stabilizers. Hence 
\begin{equation} \label{eq:logicalgroupresult}
    \mathcal L = H_1(X; \mathcal A) = \begin{cases}  0 & X = \R^2 \\ \mathcal A \oplus \mathcal A & X = T^2\end{cases} \; ,
\end{equation}
and the ground state degeneracy on the torus is $|\mathcal A| = |\det K|$. 
To obtain these results, we use the polynomial formalism for translation-invariant stabilizer codes developed by Haah \cite{Haah_2013}, which we adapt to CV degrees of freedom (real variables and rotors) in App.~\ref{app:polynomial}. 
We then apply the formalism to our model and derive Eq.~\eqref{eq:logicalgroupresult} in App.~\ref{app:application}.

\subsection{Polynomial formalism for CVs} \label{app:polynomial}
We begin with a review of the polynomial formalism for translation-invariant stabilizer codes, appropriately modified to accommodate real variable and rotor degrees of freedom. For an abelian group $A$, let $R_A = A[x^\pm, y^\pm]$ be the module of Laurent polynomials in variables $x$ and $y$ with coefficients in $A$. In the $d$-dimensional qudit setting, $A$ is typically taken to be $\Z_d$. For us, $A$ will be $\R$ when considering real variables, and $\Z$ or $\R/\Z$ when considering rotors. Periodic boundary conditions are implemented by the quotient $A[x^\pm, y^\pm] / \langle x^{L_x} = 1, y^{L_y} = 1\rangle $. We will often abuse notation and use $R_A$ for both $A[x^\pm, y^\pm]$ as well as its quotient by $x^{L_x} = 1$, $y^{L_y} = 1$, and the meaning of $R_A$ depends on whether the model is on the infinite plane or the torus.

Consider a translation-invariant system with $n_\R$ real variable degrees of freedom and $n_\Z$ rotors in each unit cell. The Pauli group of the system is 
\begin{equation}
    \mathcal P = \underbrace{R_{\R^{n_\R}} \oplus R_{\Z^{n_\Z}}}_{X \text{ type operators}} \oplus \underbrace{R_{\R^{n_\R}} \oplus R_{(\R/\Z)^{n_\Z}}}_{Z \text{ type operators}} \; .
\end{equation}
We adopt the following notation for products of Pauli operators at a single site:
\begin{equation}
    \bs{X}^{\vec{s}} = \prod_{i=1}^{n_\R} X_i^{s_i} , \quad
    \bs{Z}^{\vec{t}} = \prod_{i=1}^{n_\R} Z_i^{t_i}  ,\quad 
    \bs{\mathsf{X}}^{\vec{\mathsf{s}}} = \prod_{i=1}^{n_\Z} \mathsf{X}^{\mathsf{s}_i}_i , \quad
    \bs{\mathsf{Z}}^{\vec{\mathsf{t}}} = \prod_{i=1}^{n_\Z} \mathsf{Z}^{\mathsf{t}_i}_{i} ,
\end{equation}
for $s_i, t_i \in \R$, $\mathsf{s}_i \in \Z$, and $\mathsf{t}_i \in \R / \Z$. The $X_i, Z_i$ operators act on the real variable degrees of freedom, while the $\mathsf{X}_i, \mathsf{Z}_i$ operators act on the rotor degrees of freedom. 
A general Pauli operator is a tensor product of $\bs{X}^{\vec{s}}$, $\bs{Z}^{\vec{t}}$, $\bs{\mathsf{X}}^{\vec{\mathsf{s}}}$, $\bs{\mathsf{Z}}^{\vec{\mathsf{t}}}$ operators over different sites $\mathbf{r}= (r_x, r_y)$. Pauli operators are represented as polynomials via 
\begin{equation}
    P = \bigotimes_{\mathbf{r}}\bs{X}^{\vec{s}_\mathbf{r}}_\mathbf{r}  \bs{Z}^{\vec{t}_\mathbf{r}}_\mathbf{r} \bs{\mathsf{X}}^{\vec{\mathsf{s}}_\mathbf{r}}_\mathbf{r} \bs{\mathsf{Z}}^{\vec{\mathsf{t}}_\mathbf{r} }_\mathbf{r} 
    \longleftrightarrow 
    p = \left( 
    \begin{array}{c} 
    \sum_\mathbf{r} \vec{s}_\mathbf{r} x^{r_x} y^{r_y} \\ 
    \sum_{\mathbf{r}} \vec{\mathsf{s}}_\mathbf{r} x^{r_x} y^{r_y} \\ 
    \hline 
    \sum_{\mathbf{r}} \vec{t}_\mathbf{r} x^{r_x} y^{r_y} \\ 
    \sum_{\mathbf{r}} \vec{\mathsf{t}}_\mathbf{r} x^{r_x} y^{r_y}\end{array} \right) \; .
\end{equation}
To distinguish between Pauli operators and their polynomial representation, we use capital letters ($P$, $S$, etc.) to denote Pauli operators and lowercase letters ($p$, $s$, etc.) to refer to their polynomial representation. 
The phase obtained from commuting Pauli operators through one another is 
\begin{equation}
    P_1 P_2 = e^{2\pi i \lambda(p_1, p_2) } P_2 P_1, \quad \lambda(p_1, p_2) =  [x^0y^0]\bigl(\overline{p_1}^T \Lambda p_2\bigr) \; .
\end{equation}
Here $[x^0y^0]$ denotes the constant coefficient, taken after imposing the periodic identifications on the torus. The overline denotes the antipode map $x^{r_x} y^{r_y} \mapsto \overline{x^{r_x} y^{r_y}} = x^{-r_x} y^{-r_y}$ and $\Lambda$ is the symplectic matrix
\begin{equation}
    \Lambda = \begin{pmatrix} 0 & -\mathbbm{1}_{n \times n} \\ \mathbbm{1}_{n \times n} & 0\end{pmatrix} , \quad n = n_\R + n_\Z.
\end{equation}

A translation-invariant stabilizer code is specified by a translation-invariant, mutually commuting subgroup $\mathcal S \subset \mathcal P$ of the Pauli group that does not contain any nontrivial scalar multiples of the identity. Its code space is the mutual $+1$ eigenspace of all $S \in \mathcal S$. 
A stabilizer may be either continuous or discrete: if $S$ is a continuous stabilizer, then $S^\alpha$ for $\alpha \in \R$ are also stabilizers; if $S$ is a discrete stabilizer, then $S^k$ is a stabilizer for $k \in \Z$. 

We may pick a translation-invariant set of generators $\mathcal G$ of these stabilizers. With $g_\R$ continuous and $g_\Z$ discrete stabilizer generators per unit cell, 
\begin{equation}
    \mathcal G = R_{\R^{g_\R}} \oplus R_{\Z^{g_\Z}} .
\end{equation}
The stabilizer subgroup can be thought of as the image of a map $\sigma: \mathcal G \to \mathcal P$. 
This map sends each stabilizer generator to its corresponding Pauli operator; its kernel encodes relations between the stabilizers.
The stabilizers commute when
\begin{equation} \label{eq:sigma_commute}
    \sigma^\dagger \Lambda \sigma = 0 \; ,
\end{equation}
where the dagger operation $\dagger$ indicates transposition followed by the antipode. This condition is understood to mean exactly 0 in the block involving continuous generators, and 0 mod~1 in the block of discrete generators.

An excitation of a stabilizer code is an assignment of a phase to each stabilizer generator. The space of excitations forms $\mathcal E = \widehat{\mathcal G} = R_{\R^{g_\R}} \oplus R_{(\R/\Z)^{g_\Z}}$. There is an excitation map $\epsilon: \mathcal P \to \mathcal E$ which maps a Pauli operator to the excitations it creates, defined as 
\begin{equation}
    \epsilon = \sigma^\dagger \Lambda  .
\end{equation}
In this language, the stabilizer condition \eqref{eq:sigma_commute} can then be written as $\epsilon \sigma = 0$; any Pauli operator in the image of the stabilizer map creates no nontrivial excitations. We may summarize our discussion so far by saying that the maps $\sigma$ and $\epsilon$ form a chain complex 
\begin{equation} \label{eq:gpe}
    \mathcal G \xrightarrow{\sigma} \mathcal P \xrightarrow{\epsilon} \mathcal E.
\end{equation}
The logical operators on a stabilizer code are Pauli operators which commute with the stabilizers (and hence create no excitation), which are themselves not stabilizers. The logical group is therefore the homology of this complex at $\mathcal P$:
\begin{equation}
    \mathcal L = \frac{\ker \epsilon}{\im \sigma}.
\end{equation}

As an illustration, let us analyze the instructive example of the $\R^N$ toric code \eqref{eq:Rtoric} on the infinite plane $\R^2$ and the torus $T^2$ using this formalism. The purpose of studying this familiar example is to interpret the chain complex of \eqref{eq:gpe} as a cellular chain complex of the spatial manifold, so that computing the logical group reduces to computing cellular homology. 
This interpretation is naturally expressed in terms of dyon ribbon operators. The intuition involving ribbon operators generalizes readily to the general bosonic $K$-matrix case, which we treat in the next subsection.  

The $\R^N$ toric code has $2N$ $\R$ degrees of freedom per unit cell ($N$ on each edge), and consists of $2N$ continuous stabilizer generators per unit cell, given by the $A_{n,v}$ and $B_{n,p}$ stabilizers for $n = 1, \ldots, N$. We therefore have $\mathcal P = R_{\R^{2N}} \oplus R_{\R^{2N}}$, $\mathcal G = R_{\R^{2N}}$, and $\mathcal E = R_{\R^{2N}}$. We label plaquettes by the vertex at their southwest corner, and horizontal (vertical) edges by the vertex at their west (south) endpoint. With these conventions, the continuous stabilizers are represented in the polynomial formalism as 
\begin{equation}
    \begin{tikzpicture}[scale=1.8,baseline=-0.5ex]
        \draw (0,0)--(1,0);
        \draw (0,0)--(0,1);
        \draw (0,0)--(0,-1);
        \draw (0,0)--(-1,0);
        \node[draw=none,fill=white,inner sep=1pt] at (.5,0) {\color{blue}$Z_n$};
        \node[draw=none,fill=white,inner sep=1pt] at (0,.5) {\color{blue}$Z_n$};
        \node[draw=none,fill=white,inner sep=1pt] at (0,-.5) {\color{blue}$Z_n^{-1}$};
        \node[draw=none,fill=white,inner sep=1pt] at (-.5,0) {\color{blue}$Z_n^{-1}$};
        \node[draw=none,fill=white,inner sep=2pt] at (0,0) {$v$};
    \end{tikzpicture}\, \longleftrightarrow \begin{pmatrix} 0 \\ 0 \\ (1-\bar{x}) \bs{e}_n \\ (1-\bar{y}) \bs{e}_n \end{pmatrix}, \quad \quad
    \begin{tikzpicture}[scale=1.8,baseline=5.2ex]
        \draw (0,0)--(1,0)--(1,1)--(0,1)--(0,0);
        \node[draw=none,fill=white,inner sep=1pt] at (.5,0) {\color{purple}$X_n^{-1}$};
        \node[draw=none,fill=white,inner sep=1pt] at (1,.5) {\color{purple}$X_n^{-1}$};
        \node[draw=none,fill=white,inner sep=1pt] at (.5,1) {\color{purple}$X_n$};
        \node[draw=none,fill=white,inner sep=1pt] at (0,.5) {\color{purple}$X_n$};
        \node[draw=none,fill=white,inner sep=1pt] at (.5,.5) {$p$};
    \end{tikzpicture} \longleftrightarrow \begin{pmatrix} (y-1) \bs{e}_n \\ (1-x) \bs{e}_n \\ 0 \\ 0 \end{pmatrix} \; .
\end{equation}
Here $\bs{e}_n$ is the $N$-component unit vector with $1$ in the $n$-th component and $0$ in all others. 
The stabilizers can be organized into the stabilizer map $\sigma: \mathcal G \to \mathcal P$ taking the form 
\begin{equation}
    \sigma = \left( \begin{array}{c|c} \sigma_{A_v} & \sigma_{B_p} \end{array} \right) =  \left( \begin{array}{c|c} 0 & y-1  \\ 0 & 1-x  \\ 1-\bar{x}  & 0 \\ 1-\bar{y} & 0 \end{array} \right) \; 
\end{equation}
where each entry should be understood as a scalar multiple of the $N \times N$ identity matrix.
The excitation map $\epsilon = \sigma^\dagger \Lambda: \mathcal P \to \mathcal E$ is 
\begin{equation}
    \epsilon = \left( \begin{array}{c} \epsilon_{A_v} \\ \hline \epsilon_{B_p} \end{array} \right) = \left( \begin{array}{cccc} 1-x & 1-y & 0 & 0 \\ \hline 0 & 0  & 1-\bar{y}  & \bar{x}-1 \end{array} \right) \; .
\end{equation}

We now explain the geometric picture that will allow us to compute the logical group $\mathcal L = \ker \epsilon / \im \sigma$ as the cellular homology group of the spatial manifold $\R^2$ or $T^2$. The cellular chain groups $C_2(X; \R^{2N})$, $C_1(X; \R^{2N})$, and $C_0(X; \R^{2N})$ are formal sums with $\R^{2N}$ coefficients of faces, edges, and vertices, respectively. They can be described in the polynomial language as 
\begin{equation}
    C_2(X; \R^{2N}) = R_{\R^{2N}}, \quad C_1(X; \R^{2N}) = R_{\R^{2N}} \oplus R_{\R^{2N}}, \quad C_0(X; \R^{2N}) = R_{\R^{2N}}.
\end{equation}
With the same labeling conventions for faces and edges as before, the boundary maps $\partial_2: C_2(X; \R^{2N}) \to C_1(X; \R^{2N})$ and $\partial_1: C_1(X; \R^{2N}) \to C_0(X; \R^{2N})$ are given by
\begin{equation}
    \partial_2 = \begin{pmatrix} 1-y \\ x-1\end{pmatrix}, \quad \partial_1 = \begin{pmatrix} x-1 & y-1 \end{pmatrix}.
\end{equation}
Each entry above should be understood as a scalar multiple of the $2N \times 2N$ identity matrix. The homology group $H_1(X; \R^{2N}) = \ker \partial_1 / \im \partial_2$ vanishes for $X = \R^2$ and is $\R^{2N} \oplus \R^{2N}$ for $X = T^2$.

Notice that, as abelian groups, $\mathcal G \cong C_2(X; \R^{2N})$, $\mathcal P \cong C_1(X; \R^{2N})$, and $\mathcal E \cong C_0(X; \R^{2N})$. Importantly, these isomorphisms can be chosen so that $\sigma$ and $\epsilon$ become the boundary maps $\partial_2$ and $\partial_1$. We identify $\mathcal G \cong C_2(X; \R^{2N})$ by the map $(\bs\alpha, \bs\beta) \mapsto (\bs\alpha, \bs\beta)$ in $R_{\R^{2N}}$. This assigns the plaquette stabilizer to its corresponding face, and the vertex stabilizer to its northeast face.
We identify $\mathcal P \cong C_1(X; \R^{2N})$ using ribbon operators. 
To an edge $e$ with coefficient $(\bs{\psi}, \bs{r})$, we assign the Pauli operator
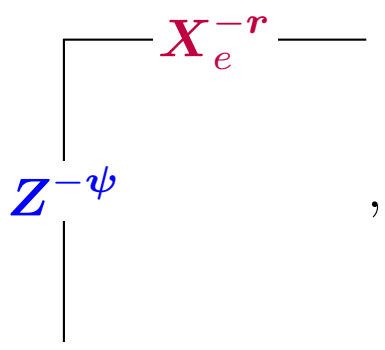
\begin{equation} \label{eq:ribboniso_pic}
    \begin{tikzpicture}[scale=2,baseline=5.9ex]
        \draw (0,0)--(0,1)--(1,1);
        \node[draw=none,fill=white,inner sep=1pt] at (0.5,1) {\color{purple}$\bs{X}_e^{-\boldsymbol{r} }$};
        \node[draw=none,fill=white,inner sep=1pt] at (0,0.5) {\color{blue}$\bs{Z}^{-\boldsymbol \psi}$};
    \end{tikzpicture},\,\quad
    
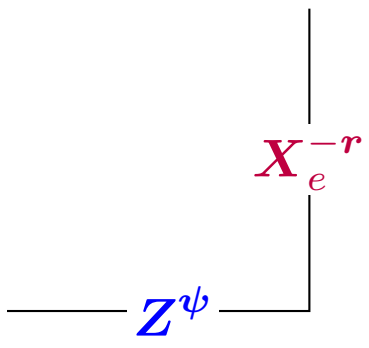
\begin{tikzpicture}[scale=2,baseline=5.9ex]
        \draw (0,0)--(1,0)--(1,1);
        \node[draw=none,fill=white,inner sep=1pt] at (.55,0) {\color{blue}$\bs{Z}^{\boldsymbol\psi}$};
        \node[draw=none,fill=white,inner sep=1pt] at (1,.5) {\color{purple}$\bs{X}_e^{-\boldsymbol r}$};
    \end{tikzpicture} \; , 
\end{equation}
which is the elementary ribbon operator $W_{a,e}$ (see Eq.~\eqref{eq:deconfined-string}) for the dyon $a = (\bs{\psi}, \bs{r})$ associated to the edge $e$. It creates the dyon $-a$ at the west (south) endpoint of $e$ and $a$ at the east (north), with each flux in the face southwest of the charge. 
Every Pauli operator on the real variable degrees of freedom can be uniquely written as a product of these elementary ribbon operators, which identifies $\mathcal P$ with $1$-chains:
\begin{equation} \label{eq:ribboniso}
    \mathcal P\ni p = \left( \begin{array}{c} \bs{r}_h \\ \bs{r}_v \\ \hline \bs{\psi}_h \\ \bs{\psi}_v\end{array} \right) \longleftrightarrow \begin{pmatrix} \begin{pmatrix} - y \bs{\psi}_v \\  - \bs{r}_h  \end{pmatrix} \\ \begin{pmatrix} x \bs{\psi}_h \\ -\bs{r}_v \end{pmatrix} \end{pmatrix} \in R_{\R^{2N}} \oplus R_{\R^{2N}} \cong C_1(X; \R^{2N}) \; .
\end{equation}
Since the elementary ribbon operator excites only its endpoints, we identify $\mathcal E \cong C_0(X; \R^{2N})$ by associating the charge and flux of each endpoint excitation to its corresponding endpoint vertex. This gives the isomorphism 
\begin{equation}
    \mathcal E \ni e = \begin{pmatrix} \bs{q} \\ \bs{\phi} \end{pmatrix} \longleftrightarrow \begin{pmatrix} xy \bs{\phi} \\ \bs{q} \end{pmatrix} \in R_{\R^{2N}} \cong C_0(X; \R^{2N}) \; .
\end{equation}
Under this identification, the flux excitation at a plaquette lives at the vertex at its northeast corner.
One can check that, using the identifications above,
\begin{equation}
    \mathcal P \ni \sigma \left( \begin{array}{c} \bs{\alpha} \\ \bs{\beta} \end{array} \right) 
    %= \left( \begin{array}{c} (1-y) \bs{\beta} \\ (x-1) \bs{\beta} \\ (1-\bar{x}) \bs\alpha \\ (1-\bar{y}) \bs\alpha \end{array} \right) 
    \longleftrightarrow 
    %\begin{pmatrix} \begin{pmatrix} (1-y) \bs{\alpha} \\ (1-y) \bs\beta \end{pmatrix} \\ \begin{pmatrix} (x-1) \bs\alpha \\ (x-) \bs \beta\end{pmatrix} \end{pmatrix} = 
    \partial_2  \left( \begin{array}{c} \bs \alpha \\ \bs \beta\end{array} \right) \in C_1(X; \R^{2N})
\end{equation}
and
\begin{equation}
    \mathcal E \ni \epsilon \left( \begin{array}{c} \bs{r}_h \\ \bs{r}_v \\ \hline \bs{\psi}_h \\ \bs{\psi}_v\end{array} \right) 
    %= \left( \begin{array}{c} (1-x) \bs{r}_h  + (1-y) \bs{r}_v \\ \hline (\bar{y}-1) \bs\psi_h + (1-\bar{x}) \bs\psi_v \end{array} \right) 
    \longleftrightarrow \partial_1 \begin{pmatrix} \begin{pmatrix} -y \bs{\psi}_v \\  -\bs{r}_h  \end{pmatrix} \\ \begin{pmatrix} x \bs{\psi}_h \\ -\bs{r}_v \end{pmatrix} \end{pmatrix} \in C_0(X; \R^{2N}) \; .
\end{equation}
We therefore identify $\sigma$ with $\partial_2$, and $\epsilon$ with $\partial_1$. 

Thus the logical group is $\mathcal L = \ker \epsilon / \im \sigma = \ker \partial_1 / \im \partial_2 = H_1(X; \R^{2N})$. On the torus, the logical group is continuous, and the code space is infinite-dimensional. The number of encoded real variable degrees of freedom follows from a generalization of the usual counting argument: on an $L_x \times L_y$ torus, there are $2N L_x L_y$ $\R$ degrees of freedom, and $2NL_x L_y - 2N$ independent continuous stabilizers which each remove one $\R$ degree of freedom. Thus the $\R^N$ toric code encodes $2N$ $\R$ degrees of freedom.

\subsection{Application to the lattice model} 
\label{app:application}
We will now use the polynomial approach to analyze the Hamiltonian \eqref{eq:Htilde} in the bosonic case, and show that it realizes the topological order described by the corresponding $K$-matrix Chern-Simons theory. In particular, we show that the group of logical operators on the torus is 
\begin{equation}
    \mathcal{L} (T^2) \cong H_1(T^2; \mathcal A) = \mathcal A \oplus \mathcal A, \quad \mathcal A = \Z^{2N} / K \Z^{2N}.
\end{equation}
Our method will be similar in spirit to the one used above in the analysis of the $\R^N$ toric code: we will show that the stabilizer and excitation maps are related to boundary maps of an appropriate cellular chain complex. The additional ingredients are the discrete Gauss law stabilizers, which have the effect of confining dyons outside of $L^\vee$, and turning closed (contractible or noncontractible) loops of ribbon operators of $L$ into stabilizers.

The Hilbert space of our model consists of $N$ real variable degrees of freedom on each edge, and $2N$ rotors on each site, for a total of $4N$ CV degrees of freedom per unit cell. The Pauli group and symplectic form are  
\begin{equation}
    \mathcal P = R_{\R^{2N}} \oplus R_{\Z^{2N}} \oplus R_{\R^{2N}} \oplus R_{(\R / \Z)^{2N}}, \quad \Lambda = \begin{pmatrix} 0 & -\mathbbm{1}_{4N \times 4N} \\ \mathbbm{1}_{4N \times 4N} & 0 \end{pmatrix}.
\end{equation}
Let us also denote by $\mathcal P_{osc}$ the subgroup of Pauli operators which only act on the $\R$ degrees of freedom
\begin{equation}
    \mathcal P_{osc} = R_{\R^{2N}} \oplus R_{\R^{2N}} \; .
\end{equation}
The Hamiltonian $\tilde{H}$ consists of $6N$ stabilizer generators per unit cell: $N$ star terms, $N$ plaquette terms, $2N$ horizontal Gauss laws, and $2N$ vertical Gauss laws. The star and plaquette stabilizers are continuous stabilizers, while the Gauss law stabilizers are discrete stabilizers. The space of stabilizer generators is therefore
\begin{equation}
    \mathcal G = \mathcal{G}_{AB} \oplus \mathcal{G}_G = (R_{\R^N} \oplus R_{\R^N}) \oplus (R_{\Z^{2N}} \oplus R_{\Z^{2N}}) \; .
\end{equation}
The space of excitations is
\begin{equation}
    \mathcal E = \widehat{\mathcal G} = \mathcal E_{AB} \oplus \mathcal E_{G} = (R_{\R^N} \oplus R_{\R^N}) \oplus (R_{(\R/\Z)^{2N}} \oplus R_{(\R/\Z)^{2N}}) \; .
\end{equation}
The stabilizer map consists of $8N$ rows and $6N$ columns, and takes the form 
\begin{equation}
\begin{aligned}
    \sigma &= 
    \left( \begin{array}{c|c|c|c} 
        \sigma_{A_v} & \sigma_{B_p} & \sigma_{G_h} & \sigma_{G_v}
    \end{array} \right)  \\ 
    &= \left( \begin{array}{c|c|c|c} 
        0 & (y-1) \mathbbm{1}_{N \times N} & \bs{Q}^T & 0 \\
        0 & (1-x) \mathbbm{1}_{N \times N} & 0 & \bs{Q}^T \\ 
        0 & 0 & (x-1) \mathbbm{1}_{2N \times 2N} & (y-1) \mathbbm{1}_{2N \times 2N} \\ 
        \hline 
        (1-\bar{x}) \mathbbm{1}_{N \times N} & 0 & 0 & -y \bs{\Phi}^T \\
        (1-\bar{y}) \mathbbm{1}_{N \times N} & 0 & x \bs\Phi^T  & 0\\
        \bs{Q} & \bs{\Phi} & -\mathcal{Q}_h & \mathcal{Q}_v
    \end{array} \right)
\end{aligned} \; .
\end{equation}
We have defined the matrices 
\begin{equation}
    \bs{Q} = \begin{pmatrix} \bs{q}_1 \\ \bs{q}_2 \\ \vdots \\\bs{q}_{2N}\end{pmatrix} , \quad 
    \bs{\Phi} = \begin{pmatrix} \bs\phi_1 \\ \bs \phi_2 \\ \vdots \\ \bs\phi_{2N} \end{pmatrix} , \quad 
    \mathcal{Q}_h =  (\bs\Phi \bs{Q}^T)_{L.T.} + x (\bs{\Phi} \bs{Q}^T)_{U.T.} , \quad
    \mathcal{Q}_v = y (\bs\Phi \bs{Q}^T)_{L.T.} + (\bs\Phi \bs{Q}^T)_{U.T.}
\end{equation}
where $L.T.$ and $U.T.$ signify the strictly lower triangular and upper triangular parts of the matrix $\bs\Phi \bs{Q}^T$. 
The excitation map $\epsilon = \sigma^\dagger \Lambda$ is 
\begin{equation}
    \epsilon = \left( \begin{array}{c} \epsilon_A \\ \epsilon_B \\   \epsilon^G_h \\ \epsilon^G_v \end{array} \right) 
    = \left( \begin{array}{cccccc} 
    (1-x)\mathbbm{1}_{N \times N} & (1-y) \mathbbm{1}_{N \times N}   & \bs{Q}^T & 0 & 0 & 0 \\
    0 & 0 & \bs{\Phi}^T & (1-\bar{y}) \mathbbm{1}_{N \times N}  & (\bar{x}-1) \mathbbm{1}_{N \times N} & 0 \\
    0 & \bar{x} \bs\Phi & -\bar{\mathcal{Q}}^T_h & -\bs{Q}  & 0 & (1-\bar{x}) \mathbbm{1}_{2N \times 2N} \\
    -\bar{y} \bs\Phi & 0 & \bar{\mathcal Q}^T_v & 0 & -\bs{Q} & (1-\bar{y}) \mathbbm{1}_{2N \times 2N}    
    \end{array} \right).
\end{equation}
It can be verified explicitly that $\epsilon \sigma = 0$, with the Gauss law excitation rows understood modulo integers. 

Our aim now is to compute the group of logical operators $\mathcal L$ on the infinite plane and the torus. Recall that $\mathcal L$ is given by the homology of the chain complex \eqref{eq:gpe} at $\mathcal P$. 
We will compute this homology group in two steps. First, we show that the support of any operator that creates no excitation (equivalently, any operator in $\ker \epsilon$) can be ``cleaned" so that it only acts on the real variable degrees of freedom and commutes with the Gauss laws. This allows us to consider instead the complex 
\begin{equation} \label{eq:reducedchaincomplex}
    \mathcal G_{osc} \xrightarrow{\sigma} \mathcal{P}_{osc} \cap \ker \epsilon_G \xrightarrow{\epsilon_{AB}} \mathcal E_{AB} \; .
\end{equation}
Here, we have defined $\mathcal G_{osc} \equiv \sigma^{-1}( \mathcal P_{osc})$ as the subspace of $\mathcal G$ which maps to stabilizers supported on the real variable degrees of freedom, and $\mathcal E_{AB}$ as the subspace of excitations which violate only the star and plaquette stabilizers.
Next, we will show that this chain complex admits a simple geometric interpretation in terms of homology on the plane and the torus. We show, for the square lattice cellulation of $X = \R^2$ and $X = T^2$, that  
\begin{equation} \label{eq:geometric_interp}
    \mathcal G_{osc} \cong C_2 (X; L^\vee) \oplus Z_1(X; L), \quad \mathcal P_{osc} \cap \ker \epsilon_G \cong C_1(X; L^\vee), \quad \mathcal E_{AB} \cong C_0(X; \R^{2N}) \; ,
\end{equation}
and provide identifications such that $\sigma$ and $\epsilon_{AB}$ can be interpreted as geometric boundary maps. It then follows that the homology of this complex is canonically isomorphic to $H_1(X; \mathcal A)$. This verifies the natural expectation that there are no nontrivial logical operators on the plane, and that the logical operators on the torus are represented by string operators supported on noncontractible cycles of the torus.

The first step is to show that any element of $\ker \epsilon$ has a representative in $\mathcal P_{osc}$. Let 
\begin{equation} \label{eq:p_parametrization}
    p = \left( \begin{array}{c} \bs{r}_h \\ \bs{r}_v \\ \bs{\mathsf{s}} \\  \hline \bs{\psi}_h \\ \bs{\psi}_v \\ \bs{\mathsf{t}} \end{array} \right), \quad 
    \bs{r}_h, \bs{r}_v, \bs{\psi}_h, \bs{\psi}_v \in R_{\R^N}, \quad 
    \bs{\mathsf{s}} \in R_{\Z^{2N}}, \quad \bs{\mathsf{t}} \in R_{(\R/\Z)^{2N}},
\end{equation}
be an arbitrary element of $\ker \epsilon$. The first $2N$ rows of $\epsilon p = 0$ are
\begin{equation}
\begin{aligned}
    (1-x) \bs{r}_h + (1-y) \bs{r}_v + \bs{Q}^T \bs{\mathsf{s}} &= 0 \\
    (1-\bar{y}) \bs{\psi}_h + (\bar{x}-1) \bs{\psi}_v + \bs{\Phi}^T  \bs{\mathsf{s}} &= 0
\end{aligned} \; .
\end{equation}
Evaluating both sides at $x=y=1$ (in other words, summing over all coefficients) gives 
\begin{equation}
    \left( \begin{array}{c} \bs{Q}^T \\  \bs\Phi^T \end{array} \right) \left. \bs{\mathsf{s}} \right|_{x=y=1} = 0.
\end{equation}
Since $\left| \det \left( \begin{array}{c} \bs{Q}^T \\  \bs\Phi^T \end{array} \right) \right| = \sqrt{|\mathcal A|} \neq 0$, we have $\left. \bs{\mathsf{s}} \right|_{x=y=1} = 0$. This means that $\bs{\mathsf{s}}$ can be written as 
\begin{equation}
    \bs{\mathsf{s}} = (1-x) {\mathsf{h}} + (1-y) {\mathsf{v}}, \quad {\mathsf{h}}, {\mathsf{v}} \in R_{\Z^{2N}} \; .
\end{equation}
We can then add Gauss law stabilizers
\begin{equation}
    p' = p + \sigma \begin{pmatrix}
        0 \\ 0 \\ {\mathsf{h}} \\ {\mathsf{v}}
    \end{pmatrix}
\end{equation}
to $p$ so that the $R_{\Z^{2N}}$ component of $p'$ vanishes. Having shown this, let us now take an arbitrary element $p \in \ker \epsilon$ with vanishing $\bs{\mathsf{s}}$, but not necessarily vanishing $\bs{\mathsf{t}}$. To this $p$, we can add $A_v$ and $B_p$ stabilizers 
\begin{equation}
    p' = p + \sigma \begin{pmatrix}
        \bs{\alpha} \\ \bs{\beta} \\ 0 \\ 0
    \end{pmatrix}, \quad \bs\alpha, \bs\beta \in R_{\R^{N}}
\end{equation}
for $\bs{\alpha}, \bs{\beta}$ satisfying
\begin{equation}
    \bs{Q} \bs{\alpha} + \bs\Phi \bs{\beta} = \begin{pmatrix} \bs{Q} & \bs{\Phi} \end{pmatrix} \begin{pmatrix}
        \bs{\alpha} \\ \bs{\beta}
    \end{pmatrix} = -\bs{\mathsf{t}}.
\end{equation}
Here, in an abuse of notation, we have used the same symbol $\bs{\mathsf{t}}$ to denote its lift into $R_{\R^{2N}}$. For a given lift, there is a unique $\bs{\alpha}, \bs{\beta}$ satisfying the above since $\left| \det \begin{pmatrix} \bs{Q} & \bs{\Phi} \end{pmatrix} \right| = \sqrt{|\mathcal A|} \neq 0$. After adding this combination of star and plaquette stabilizers, $p'$ acts nontrivially only on the real variable degrees of freedom. This allows us to restrict to the subcomplex 
\begin{equation}
    \mathcal G_{osc} \xrightarrow{\sigma} \mathcal P_{osc} \xrightarrow{\epsilon} \mathcal E \; 
\end{equation}
without changing the homology $\ker \epsilon / \im \sigma$. 
Finally, writing $\epsilon = \left( \begin{array}{c} \epsilon_{AB} \\ \hline \epsilon_G  \end{array} \right)$, we can further restrict $\mathcal P_{osc}$ to $\ker \epsilon_G$ to get \eqref{eq:reducedchaincomplex}.
Since logical operators belong to the kernel of $\epsilon$, we lose nothing by making this further restriction to $\ker \epsilon_G$. This also allows us to restrict $\mathcal E$ to $\mathcal E_{AB}$, the subspace of excitations of the vertex and plaquette stabilizers. This completes the reduction to the chain complex \eqref{eq:reducedchaincomplex}.

We have shown that the logical operators are given by the homology of \eqref{eq:reducedchaincomplex} at the middle term $\mathcal P_{osc} \cap \ker \epsilon_G$.  
We now provide the geometric interpretation of the chain complex in terms of cellular homology, beginning with the map $\epsilon_{AB}:\mathcal P_{osc} \cap \ker \epsilon_G \to \mathcal E_{AB}$. Recall from our analysis of the $\R^N$ toric code that 
\begin{equation}
    \mathcal P_{osc} \cong C_1(X; \R^{2N}) \; ,
\end{equation}
where a ribbon operator \eqref{eq:ribboniso_pic} in $\mathcal P_{osc}$ is mapped to its corresponding edge in $C_1(X; \R^{2N})$ via \eqref{eq:ribboniso}. Similarly, $\mathcal E_{AB}$ is the set of excitations of the $A_v$ and $B_p$ stabilizers. Just as in the $\R^N$ toric code, we associate to a vertex $v$ with coefficient $(\bs \phi, \bs q) \in \R^{2N}$ the dyon excitation with $\bs q$ at $v$ and $\bs \phi$ at the plaquette southwest of $v$, giving $\mathcal E_{AB} \cong C_0(X; \R^{2N})$.

Restricting $\mathcal P_{osc}$ to $\mathcal P_{osc} \cap \ker \epsilon_G$ restricts the coefficients of $C_1(X; \R^{2N})$ to $C_1(X; L^\vee)$. To see this, let $p \in \mathcal P_{osc}$ be given by \eqref{eq:p_parametrization}, with $\bs{\mathsf{s}} = \bs{\mathsf{t}} = 0$. 
Elements of $\mathcal P_{osc} \cap \ker \epsilon_G$ obey 
\begin{equation}
\begin{aligned}
    \bar{x} \bs{\Phi} \bs{r}_v - \bs{Q} \bs{\psi}_h &= 0 \mod 1 \\
    -\bar{y} \bs{\Phi} \bs{r}_h - \bs{Q} \bs{\psi}_v &= 0 \mod 1
\end{aligned} \; .
\end{equation}
These conditions imply that the ribbon operators associated to $\begin{pmatrix} -y \bs{\psi}_v \\  -\bs{r}_h  \end{pmatrix}, \begin{pmatrix} x \bs{\psi}_h \\ -\bs{r}_v \end{pmatrix}$ create dyons in $L^\vee \subset \R^{2N}$. Therefore $\mathcal P_{osc} \cap \ker \epsilon_G \cong C_1(X; L^\vee)$. 
The map $\epsilon_{AB}$ acts on $\mathcal P_{osc} \cap \ker \epsilon_G$ in precisely the same way as the excitation map $\epsilon$ of the $\R^N$ toric code. We can therefore identify $\epsilon_{AB}$ with $\partial_1$. 

Let us now understand $\mathcal G_{osc}$ and $\sigma$. First, observe that $\mathcal G$ is isomorphic to $C_2(X; \R^{2N}) \oplus C_1(X; \Z^{2N})$: given an element $g = \begin{pmatrix} \bs{\alpha} & \bs{\beta} & \bs{\mathsf h} & \bs{\mathsf v} \end{pmatrix}^T \in \mathcal G$, we can think of $\begin{pmatrix} \bs \alpha \\ \bs \beta \end{pmatrix}$ as an element of $C_2(X; \R^{2N})$ and $\begin{pmatrix} \bs{\mathsf h} \\ \bs{\mathsf v} \end{pmatrix}$ as an element of $C_1(X; \Z^{2N})$. For $g$ to be an element of $\mathcal G_{osc}$, $\sigma(g)$ must have vanishing support on the rotor degrees of freedom.  
This requirement imposes 
\begin{equation}
\begin{aligned}
    (x-1)\bs{\mathsf h} + (y-1) \bs{\mathsf v} &= 0 \\
    \bs{Q} \bs{\alpha} + \bs{\Phi} \bs{\beta} - \mathcal{Q}_h \bs{\mathsf h} + \mathcal{Q}_v \bs{\mathsf v} &= 0 \mod 1
\end{aligned} \; .
\end{equation}
The first equation above states that $\begin{pmatrix} \bs{\mathsf{h}} \\ \bs{\mathsf v} \end{pmatrix} \in C_1(X; \Z^{2N})$ is closed, so  $\begin{pmatrix} \bs{\mathsf h} \\ \bs{\mathsf v} \end{pmatrix} \in Z_1(X; \Z^{2N})$. 
To make sense of the second equation, define $\tilde{\bs{\alpha}} \equiv \bs{\alpha} - \bs\Phi^T (\bs{\mathsf v} - x \bs {\mathsf{h}})$. The second equation becomes 
\begin{equation}
    \bs Q \tilde{\bs \alpha} + \bs \Phi \bs{\beta} = 0 \mod 1 \; . 
\end{equation}
This step makes use of the fact that $\begin{pmatrix}\bs{\mathsf h} \\ \bs{\mathsf v} \end{pmatrix}$ is closed. 
Therefore, $\begin{pmatrix} \tilde{\bs{\alpha}} \\ \bs{\beta} \end{pmatrix}$ is an element of $C_2(X; L^\vee) \subset C_2(X; \R^{2N})$. This gives the isomorphism 
\begin{equation}
    \mathcal G_{osc}\ni \begin{pmatrix} \bs \alpha \\ \bs \beta \\ \bs{\mathsf{h}} \\ \bs{\mathsf{v}}\end{pmatrix}
    \longleftrightarrow 
    \left( \begin{pmatrix} \bs \alpha - \bs \Phi^T (\bs{\mathsf v} - x \bs{\mathsf h}) \\ \bs \beta \end{pmatrix} , \begin{pmatrix} \bs{\mathsf{h}} \\ \bs{\mathsf v}\end{pmatrix} \right) \in C_2(X; L^\vee) \oplus Z_1(X; \Z^{2N}).
\end{equation}
between $\mathcal G_{osc}$ and $C_2(X; L^\vee) \oplus Z_1(X; \Z^{2N})$. Under this isomorphism, the map $\sigma$ acts on $\mathcal G_{osc}$ as 
\begin{equation}
    \mathcal P_{osc} \ni \sigma \begin{pmatrix} \bs \alpha \\ \bs \beta \\ \bs{\mathsf{h}} \\ \bs{\mathsf{v}}\end{pmatrix} 
    \longleftrightarrow 
    \partial_2 \begin{pmatrix} \bs \alpha - \bs \Phi^T (\bs{\mathsf v} - x \bs{\mathsf h}) \\ \bs \beta \end{pmatrix} - \begin{pmatrix} M \bs{\mathsf h} \\ M \bs{\mathsf v}\end{pmatrix} \in C_1(X; L^\vee)
\end{equation}
where $M = \begin{pmatrix} \bs{\Phi}^T \\ \bs{Q}^T\end{pmatrix}$ is the matrix of condensed dyons. Therefore $\sigma$ acts via the boundary map on the $C_2(X; L^\vee)$ summand, and by the map of coefficients $\Z^{2N} \xrightarrow{-M} L$ on the $Z_1(X; \Z^{2N})$ summand. We can now identify our desired homology as 
\begin{equation}
    \mathcal L = \frac{\ker \epsilon_{AB}}{\im \sigma} = \frac{\ker \partial_1}{\im \partial_2 + Z_1(X; L)} =  H_1(X; \mathcal{A}).
\end{equation}
We therefore recover the fact that the logical group is trivial on the infinite plane, and $\mathcal A \oplus \mathcal A$ on the torus. The logical operators are given by the closed noncontractible ribbon operators for dyons in $L^\vee$.

It follows the ground state degeneracy on the torus is $|\mathcal A|$. This is due to the following fact:
for a GKP code on a finite number of modes with $\mathcal L$ finite, the dimension of the code space is given by $\mathrm{GSD} = \sqrt{|\mathcal L|}$. This is a standard fact in the theory of lattice GKP codes \cite{Gottesman2001Oscillator, ChakrabortyAlbert2026}, which consist of only discrete stabilizers; it also holds in our context, where some of the stabilizers are continuous.
To apply this argument, we first embed each rotor in a real variable degree of freedom stabilized by $Z=e^{2\pi i\hat x}=1$.
The key point is that, since all of the continuous stabilizer commute, and each independent continuous stabilizer removes one mode, they can be completely disentangled via Gaussian unitaries without affecting the structure of the logical group. 
Since $\mathcal L$ is finite, the discrete stabilizers act on the remaining modes as a full-rank lattice (otherwise there would be a continuous family of logical operators). 
Thus the code on the remaining modes is an ordinary multimode GKP code, which satisfies $\mathrm{GSD} = \sqrt{|\mathcal L|}$.

\end{document}